\documentclass[fleqn, usenatbib]{mnras}
\usepackage{newtxtext,newtxmath}
\usepackage[T1]{fontenc}

\usepackage{graphicx}	% Including figure files
\usepackage{amsmath}	% Advanced maths commands
\usepackage{float}
\usepackage{mathrsfs}
\usepackage{color}
\usepackage{setspace}
\usepackage{comment}
\usepackage{stfloats}
\usepackage{multirow}

\title[GaSNet-III]{Galaxy Spectra Networks (GaSNet). III. Generative pre-trained network for spectrum reconstruction, redshift estimate and anomaly detection}

\author[F. Zhong et al.]{Fucheng Zhong$^{1,2}\thanks{E-mail: zhongfch@mail2.sysu.edu.cn}$, 
Nicola R. Napolitano$^{2,3}$\thanks{E-mail: nicolarosario.napolitano@unina.it},
Caroline Heneka$^{4}$,
Jens-Kristian Krogager$^{5}$,
Ricardo Demarco$^{6}$,
\newauthor{Nicolas F. Bouch\'e$^{7}$,}
Jonathan Loveday$^{8}$,
Alexander Fritz$^{9}$,
Aur\'elien Verdier$^{10}$,
Boudewijn F. Roukema$^{11,12}$,
\newauthor{Crist\'obal Sif\'on$^{13}$},
Letizia P. Cassar\'a$^{14}$,
Roberto J. Assef$^{15}$,
Steve Ardern$^{16}$
\\
$^{1}$School of Physics and Astronomy, Sun Yat-sen University, Zhuhai Campus, 519802, Zhuhai, P. R. China \\ 
$^{2}$Department of Physics E. Pancini, University Federico II, Via Cinthia 6, 80126-I, Naples, Italy \\
$^{3}$INAF – Osservatorio Astronomico di Capodimonte, Salita Moiariello 16, I-80131 Napoli, Italy\\
$^{4}$Institut für Theoretische Physik, Universität Heidelberg, Philosophenweg 16, 69120 Heidelberg, Germany\\
$^{5}$Centre de Recherche Astrophysique de Lyon, 9 Avenue Charles André, 69230, Saint-Genis-Laval, France \\
$^{6}$Institute of Astrophysics, Facultad de Ciencias Exactas, Universidad Andr\'es Bello, Sede Concepci\'on, Talcahuano, Chile \\
$^{7}$Centre National de la Recherche Scientifique (CNRS), Centre of Research in Astrophysics of Lyon (CRAL), 9 av Charles André, F-69230 Saint Genis Laval, France \\
$^{8}$Astronomy Centre, University of Sussex, Falmer, Brighton, BN1 9QH, UK \\
$^{9}$Kuffner Observatory, Johann-Staud-Strasse 10, 1160 Vienna, Austria\\
$^{10}$Institute of Physics, Laboratory of Astrophysics, École Polytechnique Fédérale de Lausanne (EPFL), Observatoire de Sauverny, CH-1290 Versoix, Switzerland \\
$^{11}$Institute of Astronomy, Faculty of Physics, Astronomy and Informatics, Nicolaus Copernicus University, Grudziadzka 5, 87-100 Toru\'n, Poland \\
$^{12}$Univ Lyon, Ens de Lyon, Univ Lyon1, CNRS, Centre de Recherche Astrophysique de Lyon UMR5574, F--69007, Lyon, France \\
$^{13}$Instituto de F\'isica, Pontificia Universidad Cat\'olica de Valpara\'iso, Casilla 4059, Valpara\'iso, Chile \\
$^{14}$INAF-IASF Milano, via A. Corti 12, 20133, Italy \\
$^{15}$Instituto de Estudios Astrof\'isicos, Facultad de Ingenier\'ia y Ciencias, Universidad Diego Portales, Av. Ej\'ercito Libertador 441, Santiago, Chile \\
$^{16}$Department of Physics, University of Bath, Bath BA2 7AY, United Kingdom
}

\date{Accepted XXX. Received YYY; in original form ZZZ }
\pubyear{2023}

\begin{document}
\label{firstpage}
\pagerange{\pageref{firstpage}--\pageref{lastpage}}
\maketitle

\begin{abstract}
    Classification of spectra (1) and anomaly detection (2) are fundamental steps to guarantee the highest accuracy in redshift measurements (3) in modern all-sky spectroscopic surveys. We introduce a new Galaxy Spectra Neural Network (GaSNet-III) model that takes advantage of generative neural networks to perform these three tasks at once with very high
    %extraordinary 
    efficiency.
    We use two different generative networks, an autoencoder-like network and U-Net, to reconstruct the rest-frame spectrum (after redshifting). The autoencoder-like network operates similarly to the classical PCA, learning templates (eigenspectra) from the training set and returning modeling parameters. The U-Net, in contrast, functions as an end-to-end model and shows an advantage in noise reduction. By reconstructing spectra, we can achieve classification, redshift estimation, and anomaly detection in the same framework.
    Each rest-frame reconstructed spectrum is extended to the UV and a small part of the infrared (covering the blueshift of stars). Owing to the high computational efficiency of deep learning, we scan the chi-squared value for the entire type and redshift space and find the best-fitting point.
    Our results show that generative networks can achieve accuracy comparable to the classical PCA methods in spectral modeling with higher efficiency, especially achieving an average of $>98\%$ classification across all classes ($>99.9\%$ for star), and $>99\%$ (stars), $>98\%$ (galaxies) and $>93\%$ (quasars) redshift accuracy under cosmology research requirements.
    %We find that the majority of spectra can be represented with six parameters: a quadratic polynomial and three rest-frame eigenspectra.
    By comparing different peaks of chi-squared curves, we define the ``robustness'' in the scanned space, offering a method to identify potential ``anomalous'' spectra.
    Our approach provides an accurate and high-efficiency spectrum modeling tool for handling the vast data volumes from future spectroscopic sky surveys.
\end{abstract}


%%%%%%%%%%%%%%%%% BODY OF PAPER %%%%%%%%%%%%%%%%%%
\section{Introduction}
%Let's start with a famous quote from Richard Feynman, “What I cannot create, I do not understand,” which is the central idea of the whole article. 

Spectroscopic surveys, complemented with their accurate redshift measurements, map the three-dimensional matter field of the local and far universe. Among the primary scientific purposes of these surveys is to constrain the cosmological structure based on the statistics of galaxies (e.g., \citealt{2020MNRAS.499..210N, 2020MNRAS.498.2492G, 2021MNRAS.501.5616D}). Meanwhile, it also provides comprehensive spectral data products of extragalactic and galactic targets for specific research purposes.
One excellent example is the SDSS survey \citep{2003AJ....126.2081A}, which has observed more than a million galaxies, stars, and quasars (QSOs). The multi-fiber spectrograph enables the efficient observation of a large number of different targets within a specified field of the sky, with 1,000 fibers used in the SDSS-III facilities. The upcoming/ongoing stage-IV spectroscopy surveys, 4MOST \citep{2019Msngr.175....3D} and DESI \citep{2022AJ....164..207D} will be even more data-intensive, with tens of millions of spectra expected to be produced in future observations. For instance, 4MOST telescope facilities have more than 1,500 fibers, and their resolution will be larger than 5,000 for the low-resolution spectrum, covering over 15,000 deg$^2$ and producing over 20 million spectra in a five-year survey \citep{2012SPIE.8446E..0TD}.

The vast volume of data products practically makes sky surveys a playground for machine learning in data analysis and mining. In fact, this is already happening. Machine learning has become increasingly common in astronomy in recent years \citep{2019arXiv190407248B}. In the field of optical spectra, various deep learning (DL)-based methods are already being used for parameter estimation and classification. For instance, \citet[VIPERS]{2013A&A...557A..16M, 2013MNRAS.428.1424M} applied a support vector machine for the classification of galaxies, stars, and AGNs between $0.4 < z < 1.2$, and used automatic PCA for classifying spectra and repairing parts of the spectra affected by gaps, noise, and sky residuals.
\citet[StarNet]{2018MNRAS.475.2978F} and \citet{2023arXiv230605086G} have applied DL to estimate stellar properties, and \citet{2020MNRAS.491.2481B} to estimate galaxy photometry. Additionally, \citet[astroNN]{2019MNRAS.483.3255L} and \citet[SPCANet]{2020ApJ...891...23W} used DL to predict multi-element abundances. Furthermore, \citet{2019MNRAS.482..313L} employed DL for detecting strong gravitational lenses. Other studies, such as \citet{2006ApJ...650..497B, 2017MNRAS.464.4463K, 2017MNRAS.465.4311W, 2020PASP..132d4503Z, 2024MNRAS.527.1163W}, utilized DL in classification. \citet{2018arXiv180809955B} and \citet[QuasarNET]{2022MNRAS.511.4490R} applied DL for QSO classification and redshift estimation to the eBOSS DR16Q catalog \citep{2020ApJS..250....8L}, which includes the visually inspected QSOs. In this context, we have developed a series of tools for detecting emission lines from strong gravitationally lensed background galaxies embedded in the spectra of foreground lens systems \citep[GasNet-I]{2022RAA....22f5014Z}, and for detailed classification and redshift prediction with the corresponding errors \citep[GaSNet-II]{2024MNRAS.532..643Z}.

All these methods typically utilize supervised learning, relying on labeled training data to infer the relationship between input and labels. On the other hand, traditional methods typically look for the best-fit values by modeling the spectrum, e.g., through template fitting or principal component analysis (PCA; e.g., \citealt{2004AJ....128..585Y, 2004AJ....128.2603Y, 2012AJ....144..144B, 2013MNRAS.428.1424M}).
Modeling the spectrum provides greater interpretability for predictions, which is scientifically expected. 
%Reconstruction in DL is similar to modeling in traditional methods. Hence, we can expect the reconstruction to achieve the same level of predictive accuracy. 
Reconstruction in DL is similar to modeling in traditional methods and may achieve the same level of predictive accuracy.
To assess this, we decided to use two different generative neural networks to reconstruct the rest-frame spectrum: an autoencoder-like network \citep{hinton2006reducing} and U-Net \citep{2015arXiv150504597R}. By comparing the input observed spectrum with the reconstructed spectrum, we can determine the best-fit type and redshift by minimizing the chi-squared value.
The idea is simple and straightforward: we determine the components or eigenspectra needed to reproduce the observed spectrum under given conditions and identify their positions in the rest-frame reconstructed spectrum. 
All of the pixels in the observed spectrum will be compared with a rest-frame spectrum constructed by the network, from the largest to the minimum wavelength of the rest-frame spectrum, and the best-fit redshift will be decided. This technique is similar to self-supervised learning using partial masks \citep[see][]{2020arXiv200608218L}, where the masks or ignored regions correspond to wavelengths outside the observed range in the rest frame.

The overwhelming volume of data from Stage-IV sky surveys will provide one of the best opportunities for data-driven science, but it also poses a challenge to data-analysis efficiency, especially in time-domain astronomy. Owing to the high efficiency of DL, we can scan the entire parameter space of types and redshifts within milliseconds for each spectrum, generating chi-squared curves for millions of spectra in one hour. By comparing the different local minima of these chi-squared curves, we can classify spectra by their nature (e.g., stars, galaxies, or active galactic nuclei or QSOs), identify anomalies, and determine the redshift of the observed objects. Thus, all these tasks can be reduced to a problem of finding and comparing the peaks of the chi-squared curves. 
Therefore, this work combines DL-based methods with traditional spectral analysis to perform a textbook case of data-driven science.

The paper is organized as follows: In Section \ref{sec: data}, we introduce the data distribution and preprocessing; in Section \ref{sec: method}, we introduce the method of the spectral modeling with generative network models; in Section \ref{sec: results}, we show the training and testing results; in Section \ref{sec: discussion}, we discuss the performance with respect to different eigenspectra and potential anomalies. In Section \ref{sec: conclusion}, we present our conclusions.

\begin{table}
    \centering
    \caption{An overview of the data structure. Subclass nan' labels the spectra with low SNR line features that cannot be subcategorized.}
    \label{table: data info}
    \begin{tabular}{ccccccc}
    \hline \hline
    Class & Subclass  & $\overline{z}$ & $\overline{SNR}$ & $\overline{\chi^2}$ & Num\\
    \hline \hline
    STAR & A0 & 0.0 & 26.2 & 1.81 & 20,000\\
    \hline
    STAR & F5 & 0.0 & 30.5 & 1.52 & 20,000\\
    \hline
    STAR & F9 & 0.0 & 34.9 & 1.53 & 20,000\\
    \hline
    STAR & G2 & 0.0 & 33.7 & 1.26 & 20,000\\
    \hline
    STAR & K1 & 0.0 & 32.8 & 1.84 & 20,000\\
    \hline
    STAR & K3 & 0.0 & 31.1 & 1.63 & 20,000\\
    \hline
    STAR & K5 & 0.0 & 31.0 & 1.79 & 20,000\\
    \hline
    GALAXY & nan & 0.46 & 5.82 & 0.96 & 19,997\\
    \hline
    GALAXY & AGN & 0.21 & 14.3 & 1.68 & 20,000\\
    \hline
    GALAXY & STARBURST & 0.15 & 9.78 & 1.96  & 20,000\\
    \hline
    GALAXY & STARFORMING & 0.11 & 12.4 & 1.13 & 20,000 \\
    \hline
    QSO & nan & 1.62 & 2.64 & 1.04 & 19735\\
    \hline
    QSO & BROADLINE & 1.78 & 6.54 & 1.58 & 19996\\
    \hline \hline
    \end{tabular}
\end{table}

\section{Data}
\label{sec: data}

\begin{figure}
    \centering
    \includegraphics[width=0.45\textwidth]{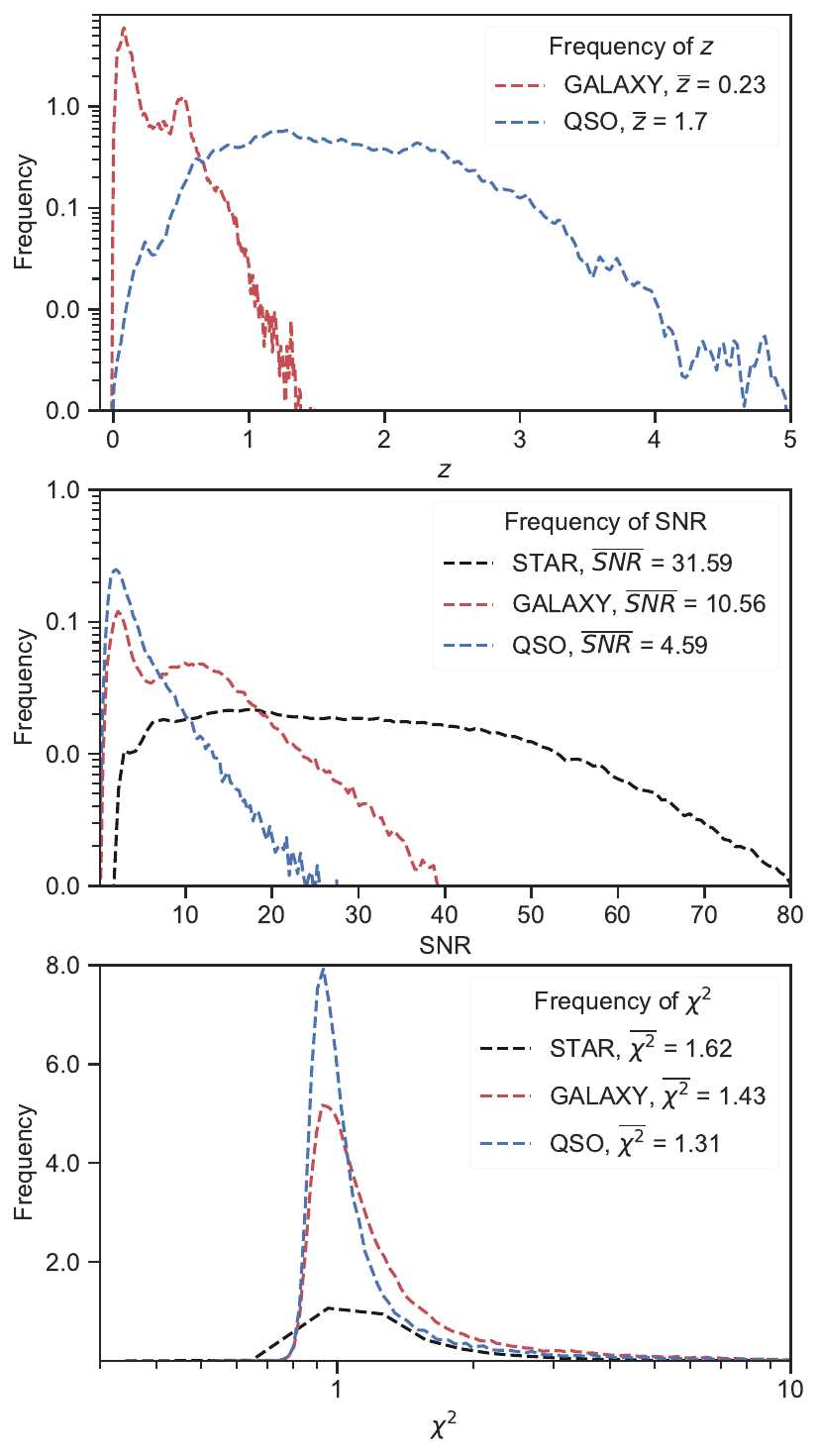}
    \caption{The density distributions of redshift, SNR, and SDSS best-fit $\chi^2$. The average values are shown in the legends. Sparse samples are found in regions with high redshift and high SNR.}
    \label{fig: data_distribution}
\end{figure}

The data used in this paper are taken from SDSS-DR16 \citep{2020ApJS..249....3A}, where 1-dimensional, stacked, and calibrated spectra are pre-classified into three main classes: GALAXY, STAR, and QSO. Within each class, the spectra are further categorized into different subclasses, but this further classification has not been used in this work (see \citealt{2024MNRAS.532..643Z}). However, to equally represent each subclass, we select an equal number of spectra from the most representative subclasses of each of the three main classes. This is shown in Table \ref{table: data info}, where we show the subclasses from which we randomly select $\sim$20,000 spectra with the quality flag $\text{ZWARNING} = 0$. 
%To reproduce the data distribution of the survey, 
There are seven subclasses in the STAR, four subclasses in GALAXY, and two subclasses in QSO. The subclasses in GALAXY and QSO are mainly distinguished by emission line features, such as the width and intensity of the lines. We note the presence of a ``nan'' subclass, which represents spectra that could not be sub-categorized due to low signal-to-noise ratio (SNR).
Due to the sparse sampling at high redshifts, we apply a redshift cut of $z < 1.5 \ (z_{\rm max} = 1.5)$ for galaxies and $z < 5 \ (z_{\rm max} = 5)$ for QSOs in our test samples. 
%The data structure is shown in Tabel \ref{table: data info}.
The mean redshift ($\bar z$), mean SNR ($\overline{SNR}$), number of spectra after applying the redshift cut (Num), and mean best-fit reduced chi-squared values are summarized in Table \ref{table: data info}.
The distribution densities of redshift, SNR, and SDSS best-fit $\chi^2$ are shown in Fig. \ref{fig: data_distribution}. The data distributions are generally unbalanced across the main parameter range. Looking at the redshift distributions, GALAXY-classified sources drop rapidly in numbers $z > 1$, while QSO drops in number density at $z > 4$.
On the other hand, the SNR for GALAXY and QSO is typically below 40 and 20, respectively, and generally lower than that of stars, which can be as high as 80. Finally, the best-fit $\chi^2$ shows peaks around 1.5 for the three classes, with a tail toward 8-10 almost similar in all classes. Note that the under-sampled redshift ranges, as well as low SNRs, may lead to reduced performance due to either poor training or low data quality.

The optical spectra wavelength range in SDSS spans approximately $3,600 - 10,000$ \AA \ with a resolution of $R\sim2,000$ \footnote{\url{https://www.sdss4.org/surveys/eboss/}}. We trimmed the spectra to a uniform wavelength grid from $3,600 - 9,200$ \AA \ due to heavy noise at the near-infrared end, and resampled it with equal spacing in logarithmic wavelength, with a spacing of $\delta_\lambda = 10^{-4}$. As a result, each trimmed spectrum contains 4,076 pixels in the logarithmic wavelength range of $\rm \log_{10}(3,600) - \log_{10}(9,200)$.

\begin{figure}
    \hspace{-0.55cm}
    \includegraphics[width=0.52\textwidth]{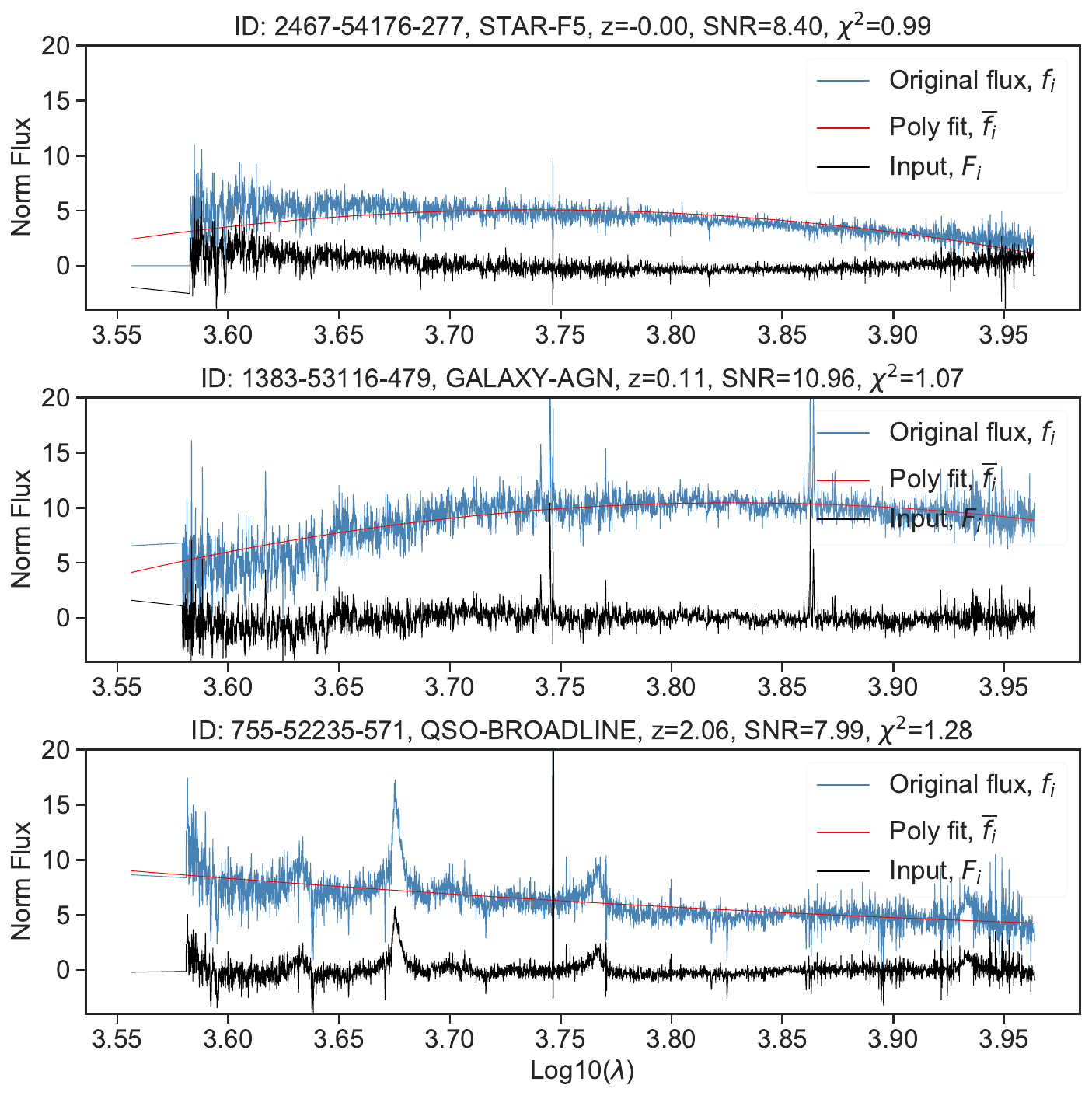}
    \caption{Examples of STAR, GALAXY, and QSO spectrum preprocessing. The ID consists of PLATE-MJD-FIBER, while class, subclass, redshift, SNR, and best-fit $\chi^2$ are shown in the panel titles.}
    \label{fig: preprocess}
\end{figure}

The spectra are preprocessed before being input into the network. First, the spectral flux is fitted with a fifth-order polynomial to model the continuum, considering the extinction or reddening \citep{2006ApJS..167...40E}. The fitted continuum is then subtracted from the flux, leaving the absorption and emission line features in the residuals. Next, the residual spectrum is normalized and made dimensionless according to the following:
\begin{align} 
    {f_i}' & = f_i - \overline{f_i}, \\
    F_i & = {f_i}'/\left(\sum_{i=1}^n {f_i}'^{2} / n\right)^{1/2}, \label{eq: preprocess}
\end{align}
where $f_i$ and $\overline{f_i}$ represent the original flux and the continuum estimated using a 5th-order polynomial fit at the $i$-th pixel, respectively, and $n=4,076$ is the number of pixels. The normalized, continuum-subtracted flux is denoted as $\mathbf F$, which will serve as the input for the network. Fig. \ref{fig: preprocess} shows some examples of spectra before and after preprocessing. Normalization removes the scale dependence of flux magnitude, while the polynomial fitting subtraction preserves the absorption and emission line features.

\section{Method}
\label{sec: method}

\subsection{Generative network as a spectrum modeling tool}

U-Net and autoencoder have usually been used in astronomy to reconstruct spectra or images. For instance, it is used in spectroscopic data reduction and denoising \citep{2020AJ....160...45P, 2021MNRAS.503.3204V, 2023MNRAS.526.3037S}, outliers detection \citep{2019BSRSL..88..174S}, features mapping \citep{2022AJ....163...71T}, and in reconstructing the rest-frame spectrum \citep{2023AJ....166...74M}. The reduced representation of a spectrum is often a latent vector of an autoencoder, which can be clustered for redshift or anomaly detection \citep{2023AJ....166...75L, 2023ApJ...956L...6L}, or even to try to derive star formation histories \citep{2024A&A...689A..58I}.
However, our approach does not directly utilize the latent vector, as it contains some stochasticity. Instead, we focus on reconstructing the rest-frame spectrum for different classes, using a specialized encoder for each class mentioned above. We then use the $\chi^2$ as a diagnostic to define the similarity between the original and reconstructed spectra and use this for classification, redshift estimation, and anomaly detection.
%and generating the chi-squared curves from multiple Encoders from different training sets 

In more detail, to reconstruct the rest-frame spectrum and calculate the $\chi^2$ at different redshifts, the reconstructed spectrum is modeled over a wavelength range extending into the ultraviolet and infrared compared to the input spectrum (which we recall is assumed to span the 3,600-9,200 \AA\ optical range). 
By comparing the input spectrum with the reconstructed spectrum and identifying the global minima on the $\chi^2$ curves, we can identify the best-fit class, redshift, and potential anomalies.
%The model spectrum is taken from the reconstruction of a generative network. After obtaining the reconstructed spectrum, we scan the entire parameters space of type and redshift, obtaining the $\chi^2$ curves. Simultaneously, we assess spectrum anomalies in this process.
In particular, focusing on the redshift, having determined the correct extragalactic class, the presence of a significant global minimum can provide a robust estimate, while the presence of multiple local minima might cause the redshift to degenerate. Typically, this is expected due to the matching pattern based on emission and absorption lines (for more detail, see Sec. \ref{sec: Model training and validation}).

%For a source with emission wavelength $\lambda_e$ at redshift $z$, the observed wavelength is $\lambda_o = \lambda_e/(1+z)$. Taking the logarithm of the wavelength results in a translation in logarithmic wavelength space:
%\begin{align} \label{eq: log(1+z)}
%    \log_{10}(1+z) = \log_{10}(\lambda_e/\lambda_o).
%\end{align}

Given that the relation between the emitted wavelength $\lambda_e$ at redshift $z$ and the observed wavelength is $\lambda_\mathrm{o} = (1+z) \lambda_\mathrm{e}$, which can be expressed in logarithmic wavelength space as $ \log_{10}(1+z) = \log_{10}(\lambda_\mathrm{o}/\lambda_\mathrm{e})$, the generative network can reconstruct the rest-frame spectrum based on two key principles:
\begin{itemize}
    \item translation symmetry with respect to $\log_{10}(1+z)$,
    \item similarity of spectra.
\end{itemize}
%We assume that spectra are sufficiently similar in the rest frame for the three classes of stars, galaxies, or QSOs, respectively.
We assume the observed spectra correspond to a finite set of rest-frame spectra that contain similar features.
%\Zhong{The corresponding rest-frame spectra contain similar features. Therefore, one might conclude that the differences in observations are from being observed at different wavelength bands after we found those typical templates.}
This implies that most observed spectra can be represented by a set of rest-frame eigenspectra. By including sufficient spectra at different redshifts in the training data set, each region of the rest-frame eigenspectra should be captured and well-modeled. The spectrum is then reconstructed in the rest frame by superimposing these eigenspectra. 

Based on this idea, we modify the autoencoder-like network architecture to learn both the eigenspectra components from the training set and the corresponding coefficients for each input spectrum, having assumed this latter to be a superposition of $m$ different rest-frame eigenspectra. 
%the encoder-only network combined with a learnable parameter matrix for predicting both the coefficients and eigenspectra. 
In this way, we can fully model the input spectrum using a DL-based method. A consequence of this architecture design is that we need to define individual networks for different classes, as each class will be characterized by a different set of eigenspectra\footnote{Technically, we can put all eigenspectra learned from different encoders trained by different classes into one encoder, and this encoder only learns the coefficients. All the different encoders can be reduced into one encoder to improve efficiency and tackle the class blending case.}. We will return to this in the next section.

Additionally, the U-Net has been proven to capture noise-complex data, e.g., in lensed galaxy image reconstruction \citep{2024arXiv241002936Z}; denoising of galaxy images \citep{2022FrASS...901043A, 2024ApJ...972...45P}; weak lensing mass map denoising \citep{2019PhRvD.100d3527S}; and gravitational wave signal denoising \citep{2023PhRvD.108d3024M}.
%We also utilize a U-Net to reconstruct spectra end-to-end. 
However, despite being very efficient for reconstruction, the U-Net does not provide the coefficients or eigenspectra needed for a more interpretative analysis. For this reason, we need to use an alternative architecture to obtain this more important information.

\subsection{Network models}
\label{sec: models}
As anticipated, two models are used for reconstructing the spectrum: the autoencoder-like model and the U-Net.
The former model (denoted as Encoder, for short) has been constructed using the encoding part of an autoencoder to map the input spectrum to $m=10$ coefficients in the final layer (see Sec. \ref{sec: num of eignespectra} for a discussion). To obtain the eigenspectra, we build a matrix with $m$ rows containing learnable parameters extracted from the same final layer above.

The reconstructed flux ($ \hat{\mathbf F}$) and the number of pixels ($N$) by the Encoder model are computed as follows:
\begin{align} 
    & (A, \omega_{m}') = \mathbf{Encoder}(\mathbf F), \\
    & \omega_{m} = {\omega_m'}^2/\sum_m({\omega_m'}^2), \\
    & \hat{\mathbf F}  = A \ \omega_{m} \cdot M_{m \times N}, \label{eq: reconstructed flux} \\
    & N = n + Log_{10}\left(\frac{1+z_{max}}{1+z_{min}} \right)/\delta_\lambda,
\end{align}
where $\mathbf F$ is the preprocessed input spectrum from Eq. \ref{eq: preprocess}, $A$ and $\omega_m$ represent the overall amplitude and the normalized coefficients, respectively, $n$ and $N$ denote the number of pixels in the input and reconstructed spectra, the parameters $z_{min}$ and $z_{max}$ are the minimum and maximum redshifts considered in the model, and $\delta_\lambda = 10^{-4}$ is the spacing of the logarithmic wavelength for the SDSS spectrum.  

\begin{figure}
    \centering
    \includegraphics[width=0.45\textwidth]{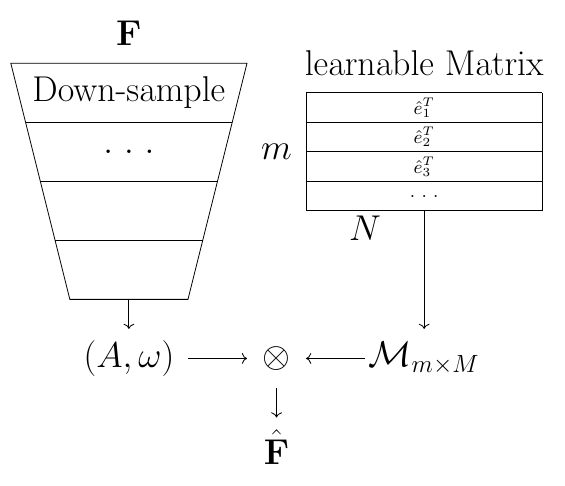}
    \includegraphics[width=0.45\textwidth]{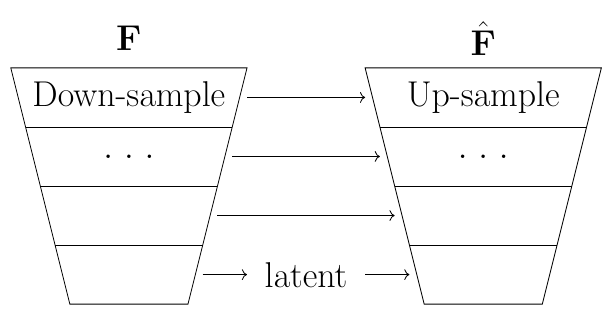}
    \caption{The network architecture. Four layers in down-sampling blocks or up-sampling blocks are used. The Encoder is illustrated in the upper panel, and the UNet is in the bottom panel. $\mathbf{F}$ represents the input spectrum with dimension $n$, while $\hat{\mathbf{F}}$ represents the output spectrum with dimension $N$, ($N > n$).
    The symbol $\otimes$ represents the multiplication of the eigenspectra by the coefficients. The eigenspectra can be extracted from the Encoder, while the UNet reconstructs the spectrum in an end-to-end way.}
    \label{fig: learnable_encoder}
\end{figure}

The coefficients are normalized to ensure their sum equals 1. The values in the eigenspectra matrix, $\mathcal{M}_{m \times N}$, are fixed once the training process is completed and become part of the model parameters, while the coefficients vary with changes in the input spectrum. The matrix $M_{m \times N}$ serves as eigenspectra, similar to the basis templates used in SDSS-III BOSS (see \citealt{2012AJ....144..144B} and \citealt{2012ApJ...753..122T}). This is shown in the upper panel of Fig. \ref{fig: learnable_encoder}, where the overall scheme of the Encoder architecture used in this paper is illustrated. The extraction of these eigenspectra and the choice of the number of eigenspectra needed for the optimal spectrum reconstruction will be discussed in detail in Sec. \ref{sec: Eigenspectra} and \ref{sec: num of eignespectra}.
Compared to a classical autoencoder, our network requires less parameters, and the eigenspectra matrix can be extracted for interpretation after training.
%The upper panel of Fig \ref{fig: learnable_encoder} illustrates the Encoder architecture used in this paper. 
This architecture is designed to identify the most common rest-frame components in the input spectra and determine their relative contributions.

If we are not interested in eigenspectra and related coefficients, the rest-frame spectrum can be alternatively reconstructed using a U-Net (UNet for short) so that we can obtain the reconstructed spectra by $\hat{\mathbf F} = \mathbf{UNet}(\mathbf F)$, where ${\mathbf F}$ represents the input spectrum, as before.
The UNet is composed of down-sampling and up-sampling blocks. The down-sampling branch consists of convolutional layers, while the up-sampling branch employs transpose convolutional layers. Skip connections between corresponding down- and up-sampling blocks allow the UNet to capture and interpret the noise in the spectrum. The UNet architecture is shown in the bottom panel of Fig. \ref{fig: learnable_encoder}.

To conclude this section, we need to remind that, for both networks, the input spectra come from different classes of objects (STAR, GALAXY, QSO). Hence, as anticipated in the previous section, we need to specialize each network to reconstruct spectra for each one of these classes using a training set selected in that specific class. A discussion about the strategy of defining this training set, i.e., whether this is to be based on theoretical models or observed data, is beyond the purpose of this paper. In our case, we use pre-classified spectra from SDSS as explained in Sec. \ref{sec: data}, which we acknowledge is prone to selection bias. 
However, depending on the class $x=(s, g, q)$ training set, then we can define a specialized Encoder$_x$ or UNet$_x$. As we will see, these will be a fundamental part of our classification step (see Sec. \ref{sec: class}).
%The training set coming from one of the fundamental steps of our process is to determine the difference between the input object and the reconstructed spectrum the network can reproduce from that input spectrum. For what we have seen in this, to optimize the reconstruction phase, we will train the two networks} 

\subsection{Loss functions and chi-squared curves}

We employed two different loss functions during the training of the Encoder and the UNet. The Encoder loss function is based on the $\chi^2$ between the reconstructed spectrum and the input spectrum, while the UNet loss function is a Smooth L1 Loss\footnote{\url{https://pytorch.org/docs/stable/generated/torch.nn.SmoothL1Loss.html}} between the input spectrum and the reconstructed spectrum, which combines the advantages of both the Mean Absolute Error (MAE) and Mean Squared Error (MSE). The Smooth L1 is used based on the assumption that it provides a noise-free representation of the input spectrum.
Since the size of the reconstructed spectrum is larger than that of the input spectrum, a proper region of the reconstructed spectrum must be selected to calculate the loss. A shift as a function of redshift, $s(z)$, is applied to align the index of the reconstructed rest-frame spectrum with that of the input observed spectrum.

The loss functions of Encoder, $L_E$, and UNet, $L_U$, are calculated as follows:
\begin{align}
%    & s(z) = -\log_{10}(z + 1)/\delta_\lambda, \\
    & L_E = \chi^2 =\frac{1}{n} \sum_{i=1}^n ( \hat F_{i+s}  - F_i)^2 / \sigma_i^2, \label{eq: L_E} \\
    & L_U = {\rm SmoothL1} = \frac{1}{n} \sum_{i=1}^n \begin{cases} 
                |\hat F_{i+s}  -  F_i| - 0.5, & \text{if } |\hat F_{i+s}  -  F_i| \geq 1 \\ 
                0.5 \ (\hat F_{i+s}  -  F_i)^2, & \text{else}
            \end{cases},
    \label{eq: L_U}
\end{align}
where $\sigma_i$ represents the error in the flux, while 
$i+s(z) = i-\log_{10}(z + 1)/\delta_\lambda$ is the rest-frame wavelength index, having resampled the wavelength range with a logarithmic step $\delta_\lambda$. The reason for adopting this formalism can be summarised as follows.
Due to the fixed wavelength range in the observed spectra, the redshift equivalently introduces a mask/window on the rest-frame spectra. From the perspective of DL, we aim to reconstruct the entire rest-frame spectra under the condition of knowing a specific unmask/window of the observed spectra. This is a common task within the DL community.
The shifting $s(z)$ represents the translation property of the spectrum as determined by the redshift. Using the loss functions defined above, we basically extrapolate the rest-frame spectrum based on information about the spectrum used in the training data.

%Once we obtain the reconstructed spectrum, the $\chi^2$ values corresponding to different redshift shifts can be calculated by using the input spectrum and the corresponding flux error to be compared with the redshifted reconstructed spectrum, with increasing shifts
Once we obtain the reconstructed spectrum, the $\chi^2$ values corresponding to different redshifts can be calculated by using the input spectrum and flux error to convolve with the reconstructed spectrum. The location of the minimum $\chi^2$ determines the best-predicted redshift.
The chi-squared curve expressed in the convolution is calculated by the following:
\begin{align}
    & \chi^2_s = \frac{1}{n-\text{dof}} \sum_{i=1}^n (\hat F_{i+s}-F_i)^2/\sigma^2_i, \\ \label{eq: chi_r^2}
    & \chi^2 = \frac{1}{n-\text{dof}} (\sigma^{-2} * \hat{\mathbf{F}}^2 - 2 (\mathbf{F} \cdot \sigma^{-2}) * \hat{\mathbf{F}} + C),
\end{align}
where $C$ is a constant for each spectrum, and the operator $*$ represents convolution\footnote{This convolution operator in deep learning differs from the usual convolution in that the kernel does not need to be flipped.}.
The $\text{dof}$ represents the degrees of freedom (in our case, with $m = 10$, $\text{dof} = 16$).
The $\chi^2_s$ represents the chi-squared value at the s-th pixel of the chi-squared curve (an s-pixel shifting in the rest frame gives one chi-squared value), while the $\chi^2$ here represents the whole chi-squared curve. The quantities $(\sigma, \hat{\mathbf{F}}, \mathbf{F})$ without the sub-index can be considered as a vector. At the same time, one can deem the $\sigma^{-2}$ and $\mathbf{F} \sigma^{-2}$ as the kernels for the convolution of the vectors $\hat{\mathbf{F}}^2$ and $\hat{\mathbf{F}}$.
The chi-squared curve is reduced as one constant term and two convolutions: one is the convolution between the inverse variance and the square of the reconstructed spectrum, while the second is the convolution between the input spectrum weighted by the inverse variance and the reconstructed spectrum.
The chi-squared curve has the same dimension as the reconstructed spectrum, $N$, indicating that we scan the redshift from $z_{min}$ to $z_{max}$.
The problem of predicting the best redshift is thus reduced to finding the minimum peak of the chi-squared curves.

After identifying the minimum chi-square and its corresponding wavelength, denoted as $\lambda_1$, the best-predicted redshift is given by:
\begin{align}
    & z_p = \lambda_o/\lambda_1 - 1, \label{eq: z_p}
\end{align}
where the $\lambda_o$ is the central wavelength of the input spectrum.
By making the derivative of the index shift $\delta s(z) = \pm 1$ (indicating one-pixel shifting), the theoretical uncertainty of the predicted redshift is:
\begin{align}\label{eq: min error}
    & \frac{\delta z_p}{1+z_p} = \mp \delta_\lambda = \mp 10^{-4}.
\end{align}

\subsection{Statistics and indicator}
\label{sec: statistical and indicator}

To evaluate the classification and redshift prediction results, we used the following quantities:
\begin{enumerate}

    \item Classification Accuracy:
    \par \begin{align}
        \text{Acc} = \frac{\text{TP}}{\text{Num}}.
    \end{align}
    The number of correct class predictions (number of true positive samples, TP) divided by the total number of test samples (Num). \\
    
    \item  The relative redshift error:
    \par \begin{align}
        \Delta z = {|z_p-z_t|}/(1+z_t),
    \end{align}
    where $z_p$ is the best-fit redshift from our model, and $z_t$ is the ground truth redshift (in our case, assumed to be the SDSS pipeline redshift). \\

    \item The mean absolute accuracy (MAE) of redshift:
    \begin{align}
        \text{MAE} = \sum_{i=1}^\text{Num} (|z_p-z_t|)/\text{Num},
    \end{align}
    where Num represents the number of spectra. \\

    \item The Median Absolute Deviation (MAD): 
    \par \begin{align}
        \text{MAD} = \text{Median} [|\Delta z - \text{Median}(\Delta z)|],
    \end{align} 
    where $\Delta z$ represents the relative redshift error. \\
    
    \item  The redshift ``robustness'':
    \par \begin{align}
        R = \chi^2_{2}/\chi^2_{1} - 1,
    \end{align}
    where $\chi^2_1$ is the global minimum peak of chi-squared curves and $\chi^2_2$ is the second-ranked minimum peak $\chi^2$. Indeed, when more than one peak with a value close to the global minimal value is found, this is the main reason for a catastrophic redshift prediction (see Sec \ref{sec: anomaly}). We call it the ``degeneracy" of redshift. Catastrophic redshifts for SDSS QSOs are identified with a deviation $\Delta z > 0.01$ \citep{2016AJ....151...44D}. Therefore, we find each local minimum peak within each $\Delta z=0.01$ interval in the $\chi^2$ curves and then used the ratio of the second minimum peak, $\chi^2_{2}$, to the global minimum value as an indicator of the best-fit redshift confidence. 
    For instance, $R=0$ means that there are at least two different $z_p$ values with the same global minimum $\chi^2$ value. In this case, models cannot give a reliable prediction, and the robustness of the predicted redshift is low. On the other hand, large $R$ values indicate more robust solutions as the second peak (if any) has a $\chi^2_{2}$ value larger than the first one ($\chi_1^2$). \\   
   
    \item The Good fraction:
    \par \begin{align}
        \text{GF}_x = \text{Num}(\Delta z < t_x)/\text{Num}.
    \end{align}
    It is defined as the ratio of the number of spectra with $\Delta z<t_x$ to the total number of test spectra.
    For the scientific requirement of cosmology measurements, $\Delta z$ should be smaller than a threshold value, $t_x$. Every measurement under that threshold is considered a good prediction. For SDSS, the requirement was $\Delta z < 10^{-3}$ for all redshifts, and $ c \Delta z < 300 + 400(z-1.5)$ for $z>1.5$ QSOs \citep{2016AJ....151...44D}. For comparison, the 4MOST requirements are $\Delta z < 10^{-4}$ for galaxies and $\Delta z < 3 \times 10^{-3}$ for QSOs.
    Due to the bin size in wavelength of the SDSS spectra and the theoretical uncertainty of our best-fit redshift (both are $\delta_\lambda = 10^{-4}$, as Eq. \ref{eq: min error} shows), we take thresholds of $t_G = 3 \times 10^{-4}$ for galaxies (which means an error within 3 pixels), $t_Q = 3 \times 10^{-3}$ for QSOs, and $t_S = 3 \times 10^{-4}$ for stars. \\  
    
    \item The Logarithmic Good fraction:
    \par \begin{align}
        & \Delta \ln(z) = \left|\ln(1+z_t)-\ln(1+z_p)\right| \\ \label{eq: Logarithmic GF}
        & \text{GF}'_x = \text{Num}\left(\Delta \ln(z) < t_x\right)/\text{Num},
    \end{align}
    where $t_x=0.0015$ for galaxies and $t_x=0.015$ for QSOs. This is introduced only to assess the performance of GaSNet-III against other standard tools in Sec. \ref{sec: comparisons}\footnote{This metric is internally used in the 4MOST collaboration for the preliminary comparison of the different pipelines.}.

\end{enumerate}

\subsection{Classification and pipeline}
\label{sec: class}
As seen in Sec. \ref{sec: models}, for each class (STAR, GALAXY, QSO) we employ a separate Encoder/UNet model to reconstruct the spectrum, assuming it comes from one of the classes in turn. We then generate the chi-squared curve corresponding to the spectrum being a STAR, GALAXY, or QSO object, respectively.

This process produces three ``chi-squared curves'' for a single input spectrum, effectively scanning the three class-type options. By identifying the minimum value among these chi-squared curves, we can pinpoint the type and redshift of the input spectrum.

\begin{figure}
    \centering
    \includegraphics[width=0.48\textwidth]{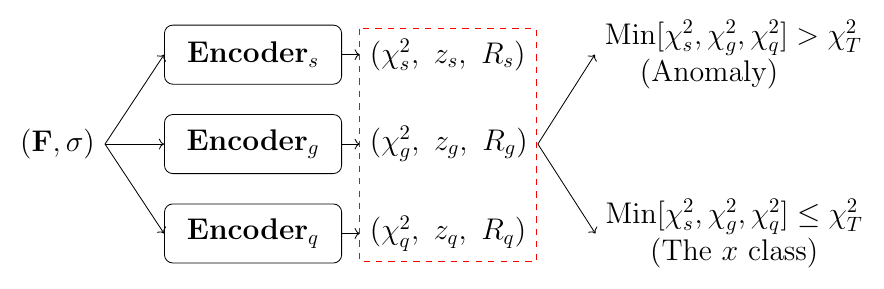}
    \caption{The pipeline for classification and anomaly detection. 
    A threshold value $\chi^2_T$ is assumed to identify outliers. Pre-trained models $\mathbf{Encoder}_s$, $\mathbf{Encoder}_g$, and $\mathbf{Encoder}_q$, which are trained on STAR, GALAXY, and QSO spectra, respectively, map the input spectrum and error, $(\mathbf{F},\ \sigma)$, to three output quantities $(\chi^2_x,\ z_x,\ R_x)$.}
    \label{fig: pipeline}
\end{figure}

For every model trained on a specific class $x=(s, g, q)$, for STAR, GALAXY, QSO, we map the input spectrum and flux error to three output quantities: $(\chi^2_x,\ z_x,\ R_x)$, where $\chi^2_x$ represents the minimum value of the chi-squared curve. The whole can be synthesized as follows:
\begin{align}
    \text{Encoder}_x(\mathbf{F},\ \sigma) = (\chi^2_x,\ z_x,\ R_x), \ \ x \in (s, g, q),
\end{align}
where $\text{Encoder}_x$ represents a model trained on the spectra of type $x$. By comparing the minimum $\chi^2$ values from different $\text{Encoder}_x$, one can obtain the classification results. The UNet is also trained in the same way to return the corresponding values $(\chi^2_x,\ z_x,\ R_x), \ \ x \in (s, g, q).$
%Except that it is used for the redshift estimate only over the class derived from the $\text{Encoder}$. 
The entire pipeline is summarized in Fig. \ref{fig: pipeline}.

\begin{figure}
    \centering
    \includegraphics[width=0.45\textwidth]{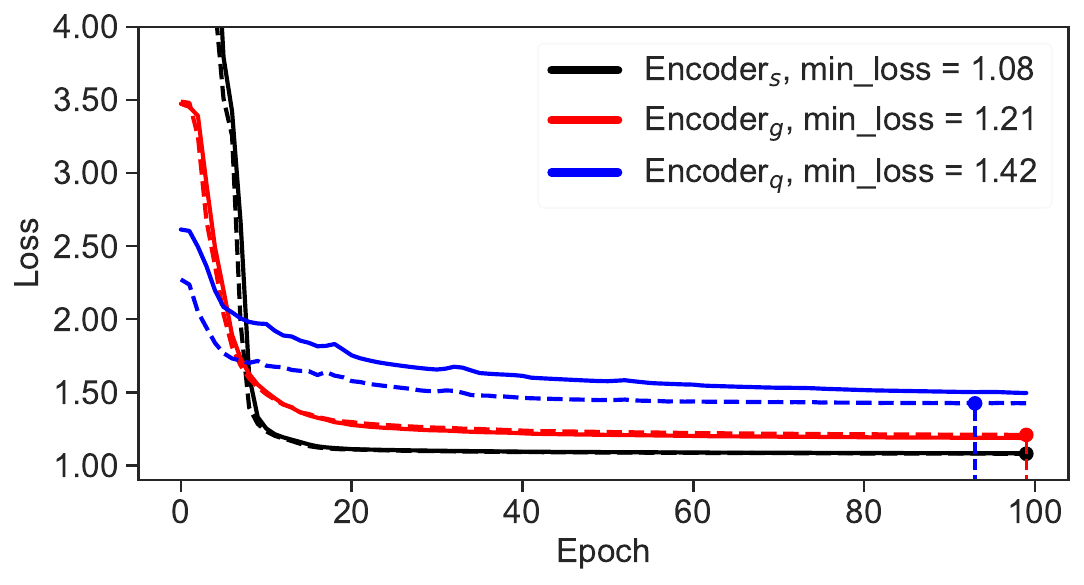}
    \includegraphics[width=0.45\textwidth]{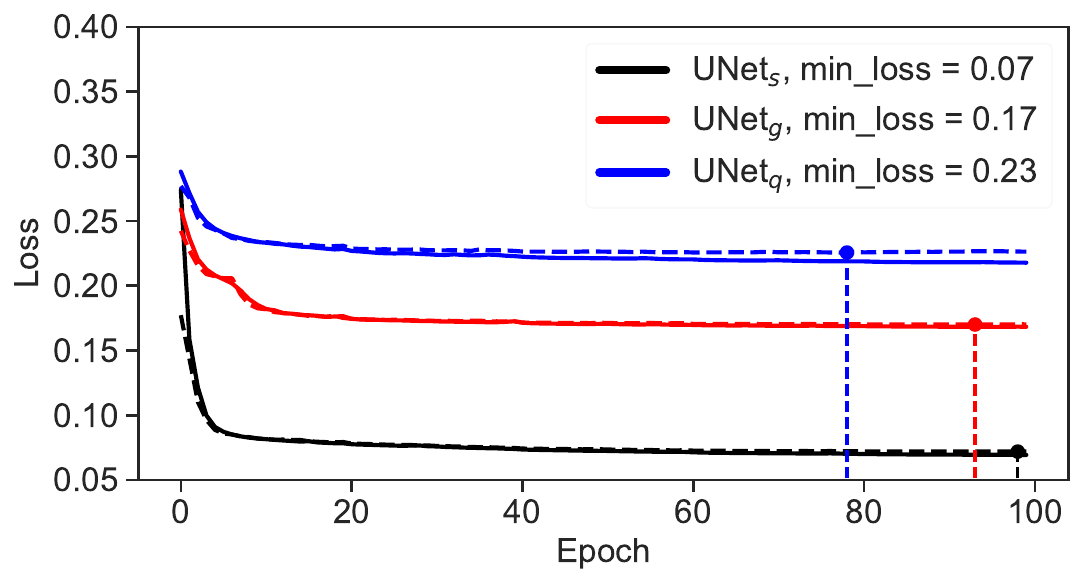}
    \caption{Loss on the training and validation in 100 epochs. The solid lines represent the training loss, while the dashed lines represent the validation loss. The minimum validation loss points are marked with vertical dashed lines, and their values are shown in the legends. The upper panel shows the loss for the three Encoders, where loss $L_E$ from Eq. \ref{eq: L_E} is used, and the lower panel shows the loss for the three UNets, where loss $L_U$ from Eq. \ref{eq: L_U} is used. The different Encoder/UNet models are trained with STAR (black), GALAXY (red), and QSO (blue) spectra, respectively.} 
    \label{fig: training loss}
\end{figure}
We finally remark that the classification step provides a way to identify anomalies based on the values of $\chi^2_x$ and the redshift robustness parameter. Because a simple definition of anomaly is difficult to give, here, we start by looking into sources that do not show any reasonable similarity with any of the classes described above. Hence, $\chi^2$-wise, this means that an anomalous candidate might be a spectrum with a large $\chi^2_x$ for every class $x=s,~g,~q$. One can expect this not to be a compelling definition, as, for instance, blending or defect spectra might fall in this simple definition. Instead, one can consider as a potential anomaly a spectrum that shows a large $\chi^2$ in one of the classes and still shows some local minima, meaning that there might be some identifiable features (absorption/emission lines) but in a very irregular pattern, such that there might be multiple solutions for a redshift (e.g., a broken continuum, artificial emission or absorption lines from bad data reduction or defects, presence of substructure in the features, etc.). More details are given in Sec. \ref{sec: anomaly}.

As a concluding remark, we note that we can further use clustering within the coefficients or latent space to perform a subclass classification or anomaly identification. This is beyond the purpose of this paper, but we give an illustration of its potential for future development in Appendix \ref{sec: outliers}.

\begin{figure*}
\hspace{-0.5cm}
    \includegraphics[width=0.75\textwidth]{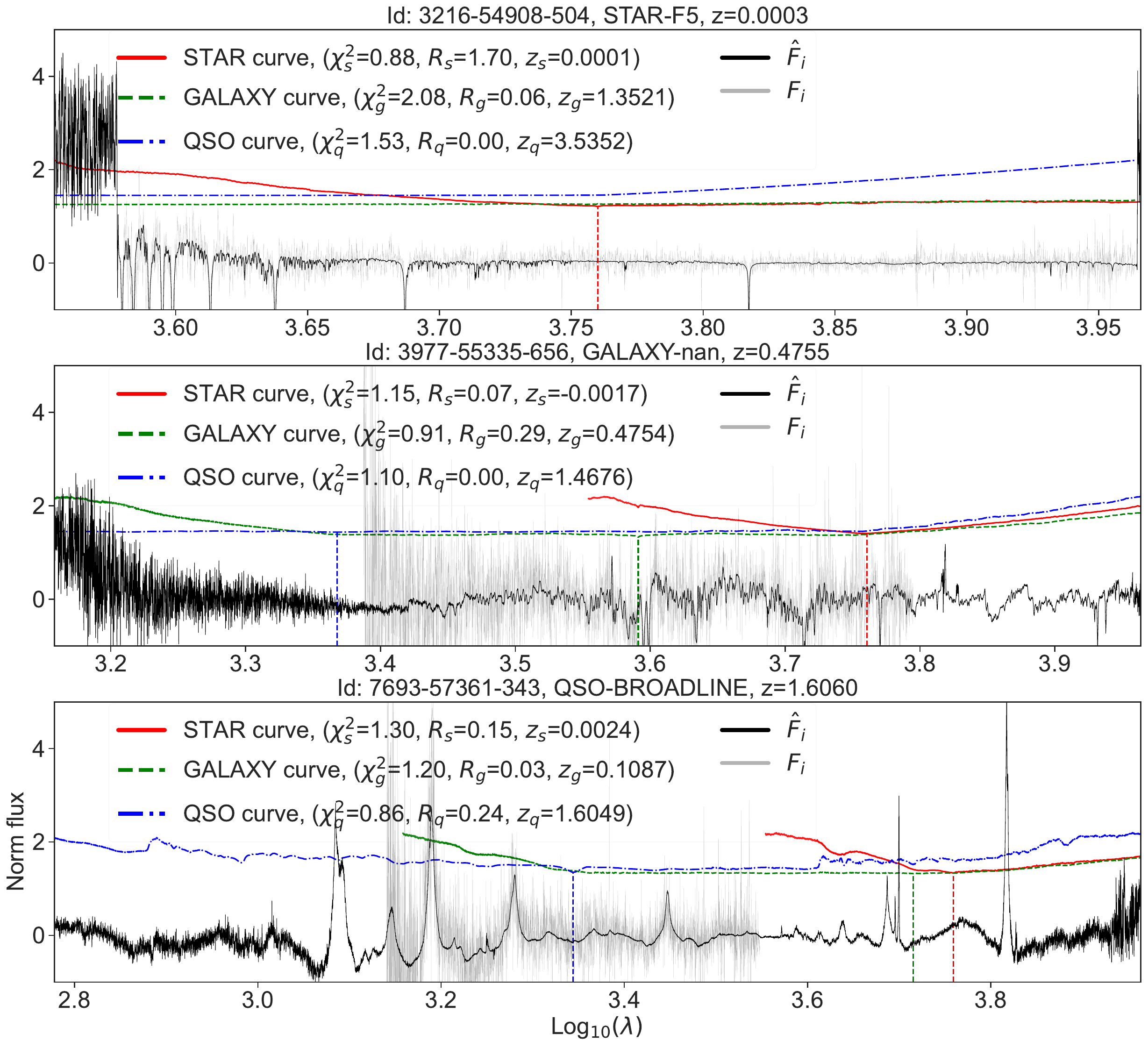}
    \caption{Example of input and reconstructed STAR/GALAXY/QSO spectra (we only show the one that has a global minimum chi-squared value in three reconstructed spectra), the chi-squared curves from three Encoders, and the corresponding output parameters in Fig. \ref{fig: pipeline}. $z$ is the SDSS pipeline redshift. The chi-squared curves have been normalized and shifted for clarity. The vertical dash line shows the global minimum $\chi^2$ for different chi-squared curves. Identifying the global minimum $\chi^2$ can estimate the type and redshift of the spectrum. The gaps at the blue and red ends of the reconstructed spectra for the STAR in the first row represent noise that is not constrained due to the lack of high Doppler-shifted stellar training spectra (lack of high redshift or blue-shift with $z \sim \pm 0.005$). For safety, the network for STAR is set to scan the range from $z = -0.005$ to $z = 0.005$.} 
    \label{fig: normal QSO}
\end{figure*}

\subsection{About the efficiency of the DL approach}

To conclude this method section, we want to address the question of whether one could use a classical approach for the spectrum reconstruction based on a given set of eigenspectra (see, e.g., SDSS templates, \citealt{2012AJ....144..144B}, or DESI templates, \citealt{2023AJ....165..144G}). In our DL approach, the Encoder will predict the corresponding coefficients once the input spectrum is given. However, based on the selected set of eigenspectra (in our cases, the ones extracted from the Encoder after training), the coefficients can also be calculated analytically. Assuming the flux $\mathbf{F}$ is a superposition of unknown coefficients $\omega$ in Eq. \ref{eq: reconstructed flux}, the $\chi^2$ curve is then calculated as shown in Eq. \ref{eq: chi_r^2}: 
\begin{equation}
\begin{aligned}
    \chi_s^2 = & \frac{1}{n} \sum_n[ \sigma^{-2}_n (\sum_m \omega_m \mathcal{M}_{m, n+s})^2 - 2 {F}_{n} \sigma^{-2}_n (\sum_m \omega_m \mathcal{M}_{m, n+s})] \\ & + C, 
\end{aligned}
\end{equation}
where $\mathcal{M}$ is the eigenspectra matrix. 
Taking the derivative of $\chi^2$ with respect to $\omega$ and setting it to zero, $\partial \chi_s^2/\partial \omega_j = 0$, we get a linear algebraic equation:
\begin{align}
    \sum_m \omega_m \sum_n \sigma^{-2}_n \mathcal{M}_{j,n+s} \mathcal{M}_{m,n+s} = \sum_n {F}_{n} \sigma^{-2}_n \mathcal{M}_{j,n+s}.
\end{align}
For each pixel point $s$ during the scan, the coefficients $\omega_{m}$ differ. By solving this equation, one can find the optimal combination of coefficients at each pixel point $s$ as:
\begin{align}
    & \omega_{m,s} = \sum_j (A_{m,j})^{-1}_s \ (B_j)_s \\
    & A_{m,j,s} = [\sigma^{-2} * (e_m^\top e_j^\top)]_s \\
    & B_{j,s} = [({F} \sigma^{-2}) * e_j^\top]_s,
\end{align}
where $e_j^\top = \mathcal{M}_{j}$, represents one row (an eigenspectrum) in the eigenspectra matrix. However, calculating $\omega_{m,s}$ analytically requires at least $m^2$ times more computational resources compared to DL predictions, in case of solely considering the computation of $A_{m,j,s}$. This demonstrates the efficiency of using a DL model for predicting these coefficients.

\section{Results}
\label{sec: results}

\subsection{Model training and validation}
\label{sec: Model training and validation}
The SDSS spectra selected for this analysis (see again Sec. \ref{sec: data}) are divided into three subsets for each class: 20,000 for model training, 5,000 for validation, and the remaining for testing (we randomly separated the STAR, GALAXY, and QSO into those three sets individually, the number of testing set is shown in the Num legend of Fig. \ref{fig: Encoder redshift}). The fixed number of training spectra serves several purposes: 
\begin{enumerate}
    \item To investigate how performance is affected by varying the amount of training data.
    \item  To demonstrate that no significant improvement occurs when the training samples reach a sufficient amount (approximately 10,000).
    \item  To test the impact of insufficient data.
\end{enumerate}
Those results will be summarized in Appendix \ref{Appendix: data number}.

For classification purposes, we have found that UNet is prone to overfitting on STAR objects, while it does not show signs of overfitting on GALAXY and QSO objects. The reason for that is that the redshift of STAR objects is close to zero, so using $F_i$ directly as training samples for UNet$_s$ causes the input and output spectra to be almost identical. In this case, learning meaningful translation symmetry and feature representations becomes difficult due to the minimum redshift. This can cause UNet$_s$ to act more like a ``copier" than a modeling tool for the spectrum (unlike the Encoder, which can reconstruct spectra using eigenspectra). This would produce an overfitting effect that would negatively impact the classification results. Indeed, since the classification relies on the global minimum chi-square, a ``copier" will always yield the minimum chi-square, regardless of the input spectra.
To prevent this, the STAR spectra are flipped by UNet$_s$ before being passed to the next downsample layer. This artificially introduced ``parity" symmetry\footnote{\url{https://pytorch.org/docs/stable/generated/torch.flip.html}} can replace the translation effects. In this way, we achieve accuracy that is very close to that of the Encoder. By introducing parity symmetry within UNet$_\mathrm{s}$, both the Encoder and UNet can serve as chi-square-generating tools in classification pipelines.

The three Encoder$_x$ and the three UNet$_x$ models ($x=s,~g,~q$) are trained using STAR, GALAXY, and QSO spectra, respectively, for 100 training epochs. A stepwise learning rate decay is applied, starting from $10^{-3}$ and gradually decreasing to $10^{-4}$. The model checkpoints corresponding to the minimum loss will be used as the final models for evaluation. Fig. \ref{fig: training loss} shows the loss of 100 training epochs. The loss drops rapidly and then decreases
slowly, converging to a stable value.

An example of STAR/GALAXY/QSO spectrum reconstruction and chi-squared curves is shown in Fig. \ref{fig: normal QSO}. Here, the input spectrum is reconstructed by three Encoders, and the corresponding chi-squared curves and parameters $(\chi^2_x, R_x, z_x)$ are shown in the legend. The reconstructed spectrum from the Encoder with the $\chi^2$ global minimum (in black) and the original input spectrum (shown in light shading) are presented and matched at the correct location (given by the best-fit redshift). The figure demonstrates that we can find the correct type and redshift by reconstructing the rest-frame spectrum and finding the global minimum of the chi-squared curves. More examples can be found in Appendix \ref{Appendix: Examples Encoder modeling}.

\subsection{Eigenspectra}

\label{sec: Eigenspectra}
\begin{figure*}
    \centering
    \includegraphics[width=0.75\textwidth]{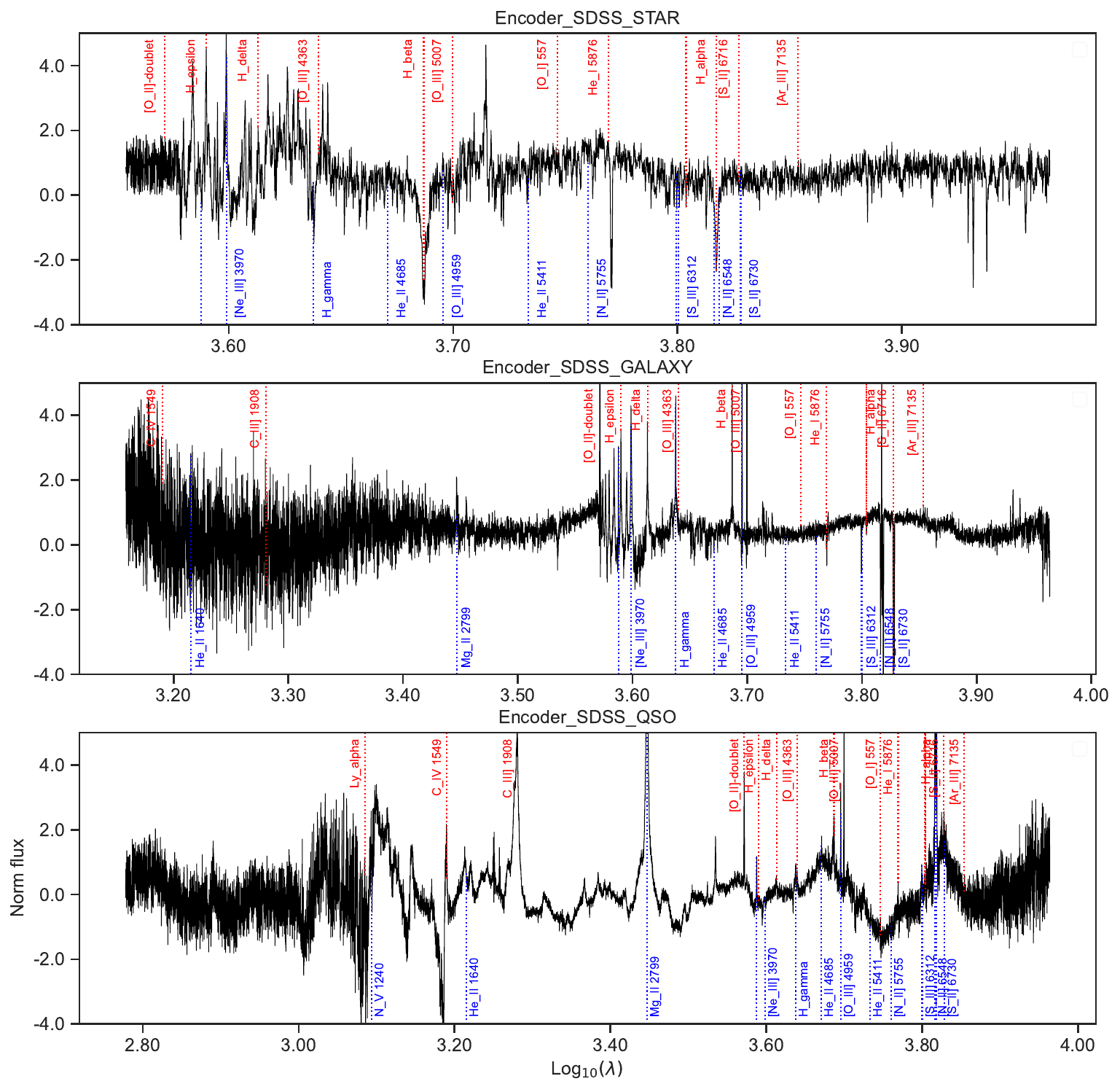}
    \caption{One of the eigenspectra learned by the Encoder. Each spectrum is just one row of the eigenspectra matrix, $\mathcal{M}$. STAR, GALAXY, and QSO eigenspectra are displayed from top to bottom. Labels in the figure indicate some relevant emission/absorption lines.} 
    \label{fig: eigenspectra}
\end{figure*}

As explained in Sec. \ref{sec: method}, the eigenspectra $\mathcal{M}$ are extracted from the Encoder model. In Fig. \ref{fig: eigenspectra}, we display the typical STAR, GALAXY, and QSO eigenspectra, with obvious emission lines features, extracted from the Encoder. In this figure, we have marked some emission lines utilized in SDSS\footnote{\url{https://classic.sdss.org/dr7/algorithms/speclinefits.php}} to demonstrate that the Encoder model effectively captures the features of emission and absorption lines, which will ultimately be used to perform the redshift predictions discussed in the next section. 

We remark that the noisy ultraviolet end of the eigenspectra shown in Fig. \ref{fig: eigenspectra} is due to the insufficient training samples at high redshifts.
%. It reflected the large uncertainties at the ultraviolet end. 
This highlights the challenge of accurately modeling spectra at higher redshifts where data sparsity affects the precision of the learned eigenspectra.

\subsection{Performance on test data}

\begin{figure*}
    \centering
    \includegraphics[width=0.8\textwidth]{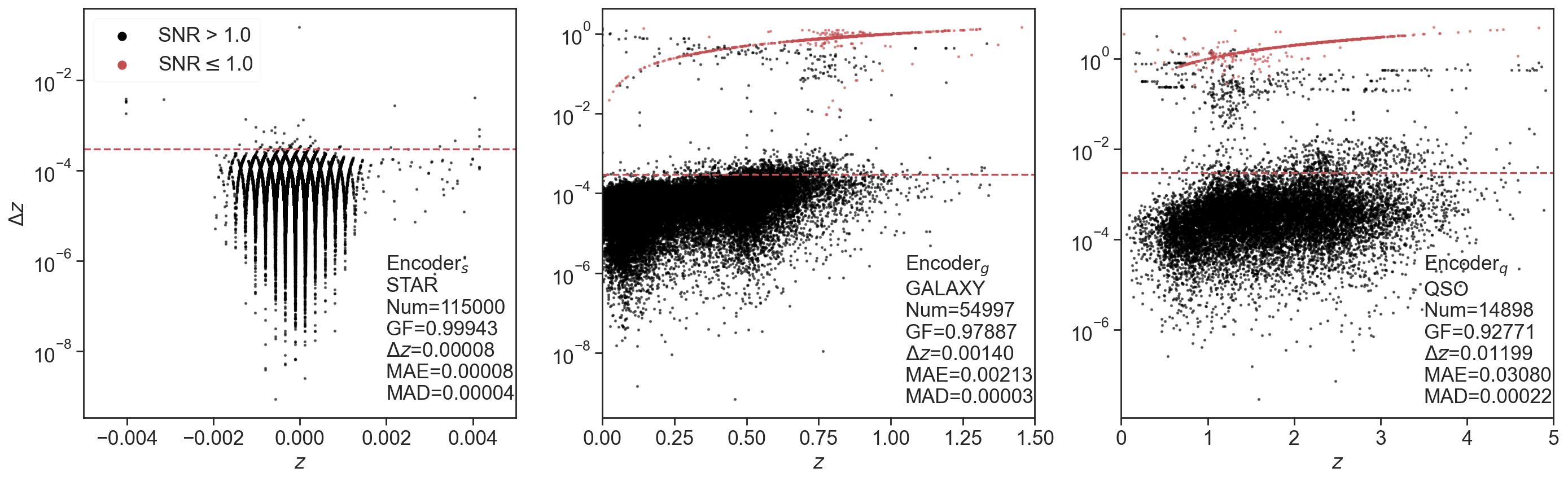}
    \includegraphics[width=0.8\textwidth]{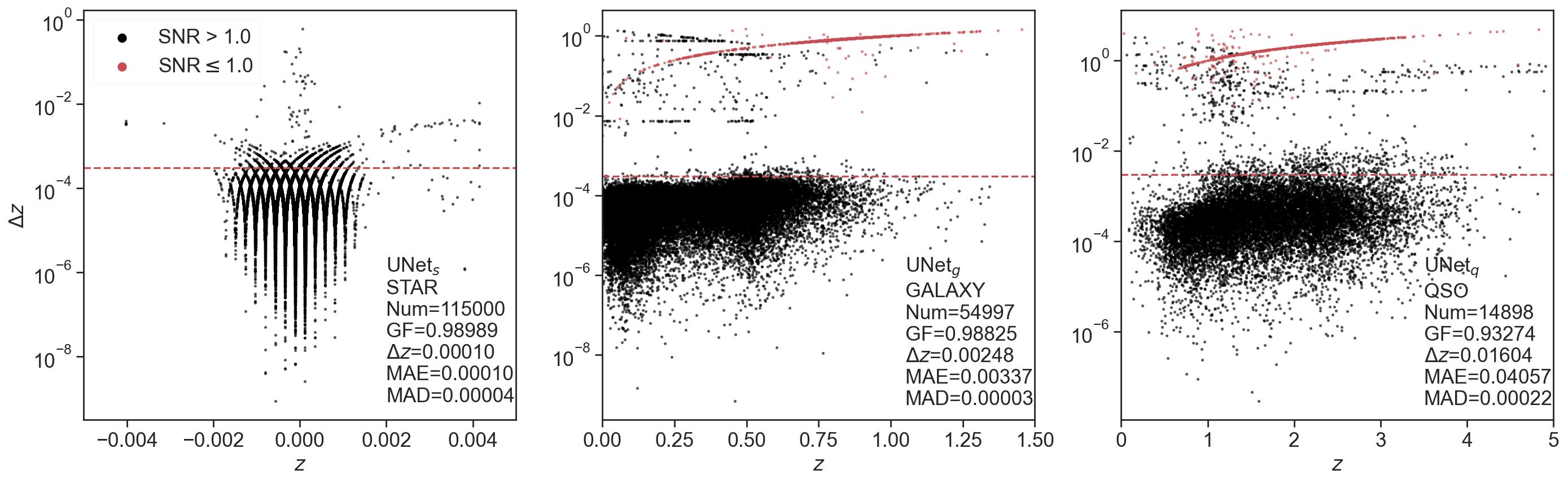}
    \caption{The redshift relative error in test data. The pattern observed in the first column for stars arises from the wavelength binning. The upper panel shows the results from Encoder models, and the bottom panel shows the results from UNet models. From the left to the right columns are STAR, GALAXY, and QSO spectra. The GF threshold, $t_x$, is drawn with a horizontal dashed line. The key statistical values are reported at the bottom right of each panel.}
    \label{fig: Encoder redshift}
\end{figure*}

The best-fit redshifts from the Encoder and UNet models applied to the test samples selected from the dataset detailed in Table \ref{table: data info} are shown in Fig. \ref{fig: Encoder redshift}. It shows the relative redshift error, $\Delta z$, vs. the SDSS redshift as compared to the cosmological survey requirement marked as a red dashed line. From the figure we can see that, except for a minor fraction of large $\Delta z$ outliers, the SDSS pipeline redshifts (which we assume as ground truth) can be accurately reproduced for most spectra. This is quantified by the GF indicator for redshift predictions within the specified threshold that is over 99\% for STAR and around 98\% for GALAXY spectra in both the Encoder and UNet models. However, for QSOs, the GF decreases to approximately 93\%, which is still an excellent performance although not as good as for the other classes. A primary reason for significant redshift deviation is that spectra with low SNR do not provide a reliable global minimum on the chi-squared curves, as indicated by the red points in the figure, where SNR $\leq 1$. 
By comparing the two networks, we see that the Encoder achieves a higher GF for STAR, which has a higher average SNR, while the UNet performs better in GALAXY and QSO spectra, which tend to have lower SNRs. Although the UNet is an end-to-end reconstruction model and does not provide the complete coefficients for modeling the spectra and corresponding eigenspectra, it demonstrates a clear advantage in capturing and reducing noise in the spectra. One can confirm this point by the chi-square comparison in Table \ref{table: chi2 and D}. In this table, we present the mean values of reduced $\chi_x$ and $R_x$ on the test data, as well as the chi-squared values for the SDSS pipeline model. The results show that the Encoder achieves a lower minimum chi-squared value for the STAR and GALAXY spectra compared to the SDSS pipeline model, although not for the QSO spectra. UNet, on the other hand, achieves a lower average minimum chi-square in all types of spectra. This indicates that our model performs at least on par with the SDSS pipeline in modeling the STAR and GALAXY spectra.

\begin{figure}
    \hspace{-0.5cm}
    \includegraphics[width=0.25\textwidth]{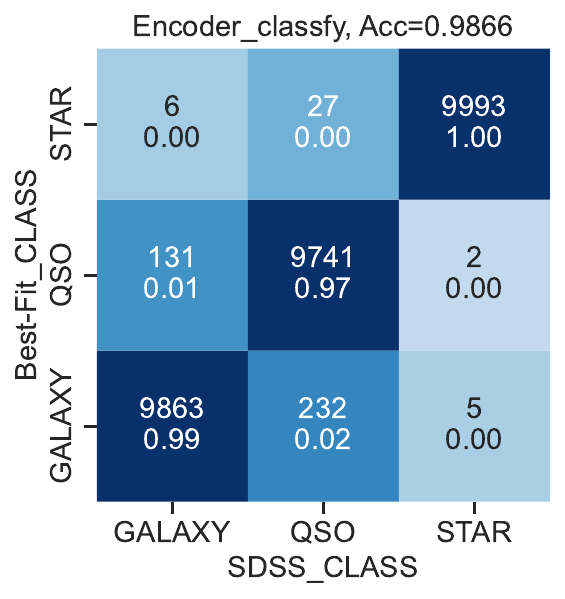}
    \includegraphics[width=0.25\textwidth]{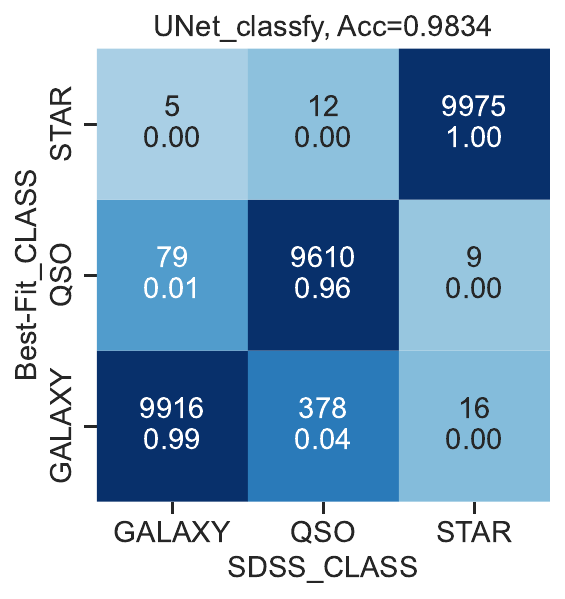}
    \caption{The confusion matrix for the classification of 30,000 test spectra. The left panel shows the results from the Encoder, while the right panel corresponds to UNet. Each matrix element contains both the number of spectra classified and the corresponding normalized value.}
    \label{fig: classify results}
\end{figure}

For the spectrum classification purpose, we randomly selected 10,000 spectra from the test set of each class to use as the test data for pipeline spectrum classification. We only considered the minimum values of $(\chi^2_s, \chi^2_g, \chi^2_q)$ to determine the best-fit type for the input spectrum. Fig. \ref{fig: classify results} presents the confusion matrix for the classification of the total 30,000 test spectra. From the metrics, we can see that the overall classification accuracy combining the three classes exceeds 98\% for both the Encoder and UNet, with the best accuracy reached for STAR types ($\sim100\%$) and slightly worse for GALAXY types ($\sim99\%$), while QSO types are the worst but with a still striking $\sim96-97\%$.
%\Zhong{In general, galaxies outnumber QSOs by a factor of approximately $\sim1,000$, meaning that even a 99\% accurate classification of galaxies could lead to contamination that completely overwhelms the number of QSOs. However, this is not the case in 4MOST, as objects have been preselected, resulting in a much higher QSO-to-galaxy ratio. By combining continuous color information with robustness parameters, we can achieve higher purity in QSO classification.}

\begin{table}
    \centering
    \caption{The average values of $\chi^2_x$ and $R_x$, and the comparison with the SDSS model.}
    \label{table: chi2 and D}
    \begin{tabular}{ccccccc}
    \hline \hline
    Model & $\chi^2_s$  & $\chi^2_g$ & $\chi^2_q$ & $R_s$ & $R_g$  & $R_q$  \\
    \hline \hline
    Encoder &  1.089 & 1.203 &  1.378 & 16.199 & 2.402 & 0.511\\
    \hline
    UNet & 1.004 & 1.163 & 1.244 & 13.766 & 3.720 & 0.607 \\
    \hline
    SDSS model & 1.624 & 1.432 & 1.304 & - & - & - \\
    \hline \hline
    \end{tabular}
\end{table}

\begin{figure}
    \centering
    \includegraphics[width=0.48\textwidth]{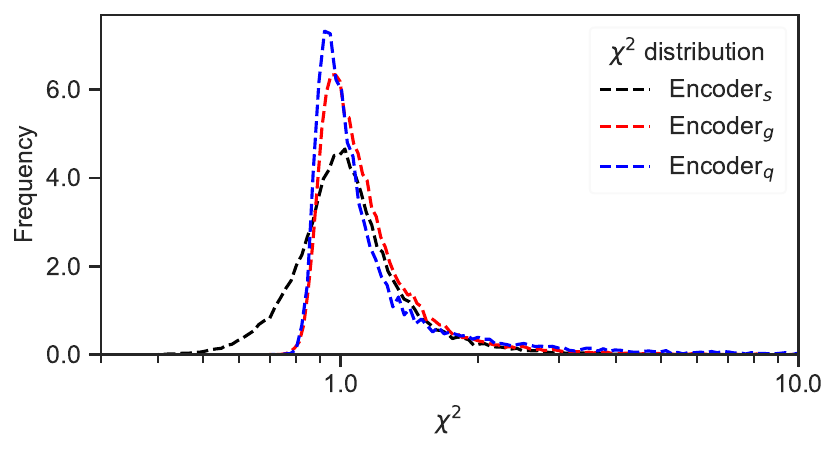}
    \caption{The minimum chi-square distribution of Encoder models.}
    \label{fig: minimum chi-square distribution}
\end{figure}

In Fig. \ref{fig: minimum chi-square distribution}, we present the minimum $\chi^2$ distribution on the test set of Encoder models. This figure shows the distribution peaks for three encoders, each approximating a value of 1, which indicates that our model can effectively reconstruct the spectrum.

\begin{comment}
\begin{figure}
    \centering
    \includegraphics[width=0.45\textwidth]{paper/fig/SNR_vs_GF.pdf}
    \caption{The variation of GF with SNR. We focus on the main SNR range (0–20) for GALAXY and QSO spectra. In this range, the SNR is divided into 20 bins. The fluctuations for QSOs at high SNR are attributed to the sparse number of samples in those bins.} 
    \label{fig: snr_vs_GF}
\end{figure}

\begin{figure}
    \centering
    \includegraphics[width=0.48\textwidth]{paper/fig/SNR_vs_Degeneracy.pdf}
    \caption{The average robustness in each bin as a function of SNR. The SNR range from 0 to 40 is divided into 40 bins.}
    \label{fig: Degeneracy vs SNR}
\end{figure}
\end{comment}

\begin{figure}
    \centering
    \includegraphics[width=0.45\textwidth]{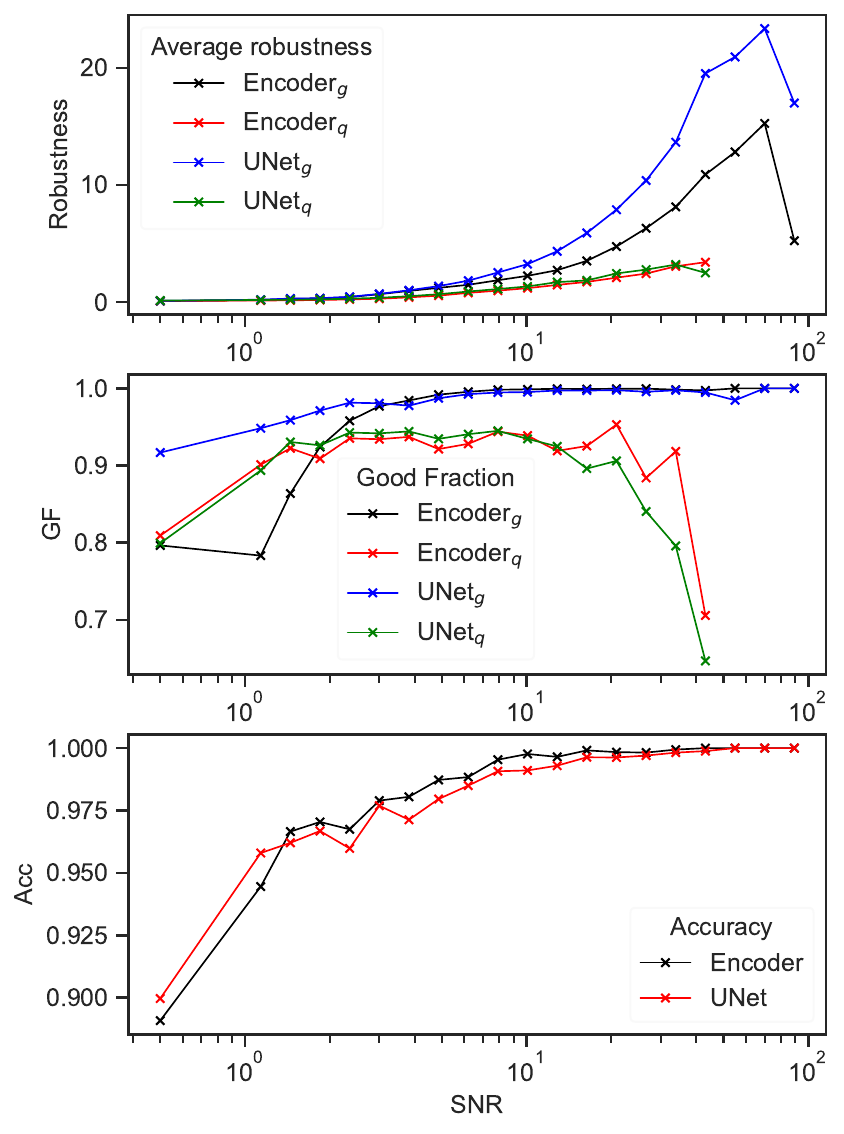}
    \caption{The variation of metrics (robustness, GF, Acc) with SNR. We focus on the main SNR range (0–100) for GALAXY and QSO spectra.
    %In this range, the SNR is divided into 20 bins 
    In the range from 1 to 100, the SNR is divided into 19 logarithmic bins, with an additional bin in the range 0-1.
    The fluctuations for QSOs at high SNR are attributed to the sparse number of samples in those bins.} 
    \label{fig: snr_vs_GF} \label{fig: Degeneracy vs SNR}
\end{figure}

\begin{figure}
    \centering
    \includegraphics[width=0.45\textwidth]{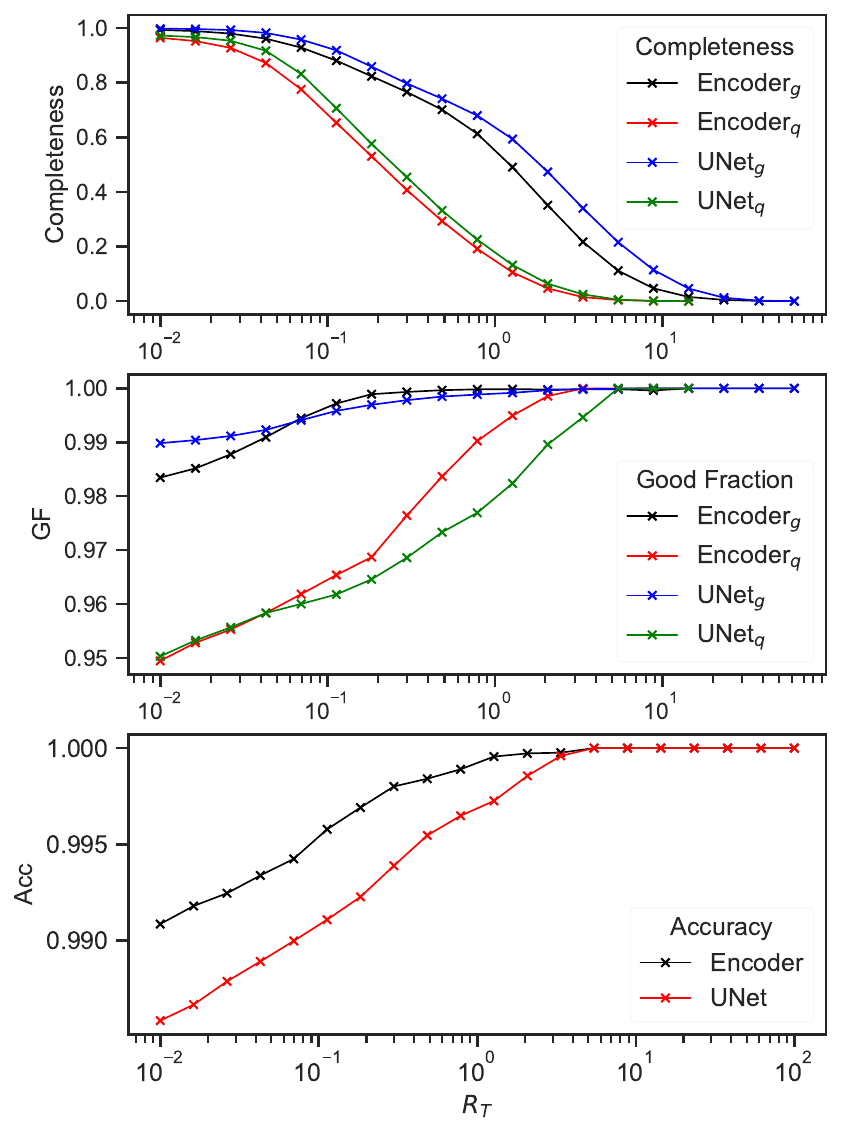}
    \caption{The variation of metrics and completeness with different $R$ threshold values ($R \ge R_T$). $R_T$ ranges from 0 to 100 and is divided into 20 bins. Taking $R_T \sim 0.1$ and $R_T \sim 1.0$, we can significantly improve the subset's GF of GALAXY and QSO.}
    \label{fig: D_vs_GF}
\end{figure}

\begin{figure}
    \centering
    \includegraphics[width=0.48\textwidth]{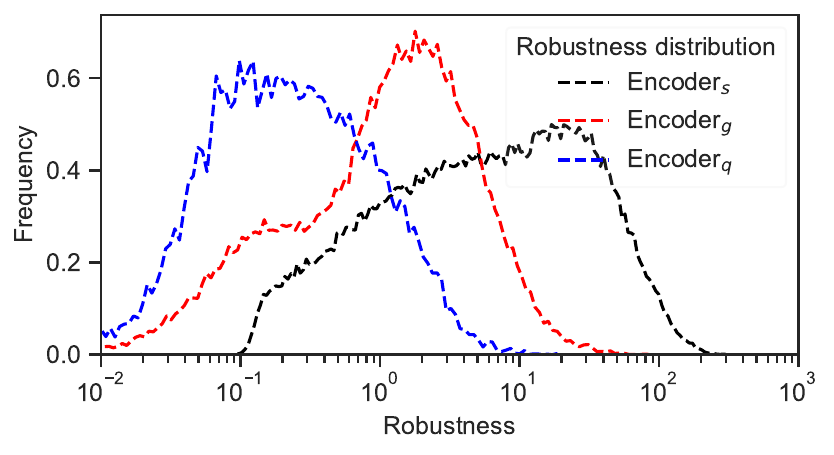}
    \caption{Robustness distribution for test set samples from the Encoders.}
    \label{fig: Degeneracy distribution}
\end{figure}

Due to the GF indicator for STAR being more than $99.9\%$, we focus only on presenting the GF of GALAXY and QSO as a function of SNR in the middle panel of Fig. \ref{fig: snr_vs_GF}.
The figure suggests that the main driver for the predicted redshift accuracy is SNR. Indeed, GF increases significantly when the SNR rises from 0 to 2, after which it stabilizes at a high value. We remark that the main impact of the SNR is on the difficulty of defining an unambiguous redshift, which is higher for lower SNRs. From this point of view, one can expect the $R$ parameter, i.e., the robustness in redshift prediction defined in Sec. \ref{sec: statistical and indicator}, to be also correlated to the SNR (see the top panel of Fig. \ref{fig: Degeneracy vs SNR}). This $R$ parameter, being connected to the ratio of the $\chi^2$ of the second-ranked minimum, provides a measure of the redshift solution quality as the larger the $R$ value, the more robust the redshift is. In fact, if there is a second local $\chi^2$ minimum, $R$ value shows how much larger the $\chi^2$ value is than the first one. In the worst-case scenario, for very low SNR (e.g., $<2$), due to noise overwhelming the spectrum, no significant emission or absorption features were identified in the spectrum, and the $\chi^2$ scanning shows smooth variation with no clear peak producing a $R\sim 0$, i.e., low robustness for the redshift solution. In general, one can define a threshold, $R_T$, to judge if a solution is robust enough, i.e., if the $\chi_2^2$ is significantly larger than $\chi_1^2$.
In Fig. \ref{fig: D_vs_GF} (middle), we report the variation of GF for different $R_T$ thresholds for the subset with $R > R_T$, where we can see that the larger the $R_T$, the larger the overall GF, with a steep increase for $R_T<1$, mirroring the SNR trend discussed above. In particular, for $R_T \sim 0.1$, the GALAXY type already has a GF$\sim 1$. The completeness of this subset is defined as the ratio of the number of spectra with $R$ greater than a given threshold to the total number of samples, $\text{Num}(R>R_T)/\text{Num}$.
In this case, in the same Fig. \ref{fig: D_vs_GF} (top), we show the completeness as a function of $R_T$, showing that GF increases significantly as the $R$ threshold rises from 0 to 0.5, while the corresponding completeness decreases as the fraction of spectra with higher robustness decreases. However, the fact that the corresponding GF is $\sim1$ means that for a large $R_T$, all GF indicator solutions are also robust. 

In the bottom panel of Fig.~\ref{fig: snr_vs_GF} and \ref{fig: D_vs_GF}, we report the variation of average accuracy for different SNR or $R_T$ subset with $R>R_T$. As expected, the accuracy shows a positive correlation with both SNR and $R_T$.
%The figure indicates that peak degeneracy is another factor that causes catastrophic predictions.
%The main reason for peak degeneracy is the low SNR. In this case, due to noise overwhelming the spectrum, no significant emission or absorption features were presented in the spectrum, and the chi-square scanning failed to identify a global minimum compared to other wavelength locations. Fig \ref{fig: Degeneracy vs SNR} shows the average robustness as the function of SNR. The robustness is positively correlated with the SNR.
In Fig. \ref{fig: Degeneracy distribution}, we finally show the robustness distribution of the Encoders, illustrating that the primary regions for the three different types of spectra fall within distinct ranges. This robustness distribution is mainly driven by the SNR differences among these spectra.
%Defining robustness helps reflect the confidence in redshift estimation and model reconstruction. A well-chosen threshold can maximize GF while maintaining relatively high completeness. 

\subsection{Comparisons}
\label{sec: comparisons}
\begin{table}
    \centering
    \caption{The comparison of different methods. The time is measured in seconds per spectrum per core (sec/spec/core). The indices in MAD$_g$/GF'$_g$ (Eq. \ref{eq: Logarithmic GF}) and MAD$_q$/GF'$_q$ represent GALAXY and QSO spectra, respectively.}
    \label{table: Comparison different methold}
    \begin{tabular}{cccccc}
    \hline \hline
    Method & MAD$_g$ & GF$'_g$ & MAD$_q$ & GF$'_q$ & Time\\
           & $/10^{-4}$ &   & $/10^{-4}$  &   &  \\
    \hline
    QXP++ & 3.7 & 0.980 & -  & - & 0.4 \\
    \hline
    Redmonster & 0.27 & 0.999 & 4.52  & 0.977  & <0.4 \\
    \hline
    pPXF  & 772.3 & 0.854  & - & - & 0.03 \\
    \hline
    Encoder & 0.33 & 0.994 & 2.04 & 0.982 & <0.001 \\
    \hline
    UNet & 0.31 & 0.994 & 2.65  & 0.954 & < 0.001 \\
    \hline \hline
    \end{tabular}
\end{table}

In this section, we compare the performances of the GaSNet-III models against three different standard software for redshift inferences: QXP++ {(Davies et al. in prep)}, Redmonster \citep{2016AJ....152..205H}, and pPXF \citep{2023MNRAS.526.3273C}. 
We used the same catalog of test samples, randomly selected from the SDSS, which included 7,886 galaxies and 7,886 QSOs. The comparison results are summarized in Table \ref{table: Comparison different methold}.
%It's important to note that the GF used in this comparison is defined differently from the one in \ref{sec: statistical and indicator}. Here, GF' is defined as:
%\begin{align}
%    & \Delta \text{log}z = |\text{log}(1+z_t)-\text{log}(1+z_p)| \\
%    & \text{GF}' = \text{Num}(\Delta \text{log}z < t_x)/\text{Num},
%\end{align}
%where $t_x=0.0015$ for galaxies and $t_x=0.015$ for QSOs.

Once again, we focus on extragalactic sources (GALAXY and QSO types) as these are the most relevant sources on which to test performances. The first obvious thing to note is that the GaSNet-III models, either Encoder or UNet, show very low MAD, both for galaxies and QSOs, which are much lower than those from some of the classical methods (e.g., pPXF for galaxies), but generally comparable to the values from Redmonster for galaxies and even better than Redmonster's values for QSOs. 
In particular, this latter result is significant as high accuracy for QSOs is usually challenging to reach due to the high degeneracies resulting from the broad width and variety of emission lines in their spectra (see, e.g., \citealt{2010MNRAS.405.2302H}). In terms of GF$'$ (Eq. \ref{eq: Logarithmic GF}), the GaSNet-III performance is statistically consistent with the best performance reached by Redmonster, except for the GF$'_q$ of the UNet, which is around 2\% lower than Redmonster's.
%The table shows that our method can reach the level of accuracy of classical methods and even surpass them in some cases. 
If we combine these excellent accuracies with the computational time, which is about three orders of magnitude faster, {\it we conclude that the DL approach is a very competitive alternative to the best-performing classical methods for redshift measurement}. In the next section, we will show that the significant advantage of this approach is the possibility of combining accurate and fast performance in redshift estimates with the ability to detect anomalies, both for quality checks and for new discoveries.
%Our model is much faster than other methods. 
%Although the QSO GF' of UNet is around 2\% lower than $\mathbf{Redmonster}$, the other statistical values are close even to represent the best performances. This demonstrates that our model is efficient in terms of both accuracy and computational speed.

\section{Discussion}
\label{sec: discussion}
In the previous section, we presented the results from the GaSNet-III models for redshift determination. Since the performances are dependent on some choices we have made for the spectrum reconstruction, in this section, we discuss the tests done in terms of the number of eigenspectra and the minimum reconstruction level to optimize the models. This will naturally bring us to the question of whether the reconstructed spectra can allow us to detect anomalies in the true spectra. ``Anomalies'' are important for two reasons: 1) they can point to a lack of information in the reconstructed spectra, or 2) more interestingly, they can highlight features that deviate from regular patterns in a given spectral class, hence opening up avenues to new discoveries.

\subsection{The minimum representation of a spectrum}
\label{sec: num of eignespectra}

\begin{table}
    \centering
    \caption{The performance of using: 1) a quadratic polynomial fitting in spectrum preprocessing, and 2) a number of $m=2,5,10$ eigenspectra for the Encoder. GF, $\Bar{\Delta} z$, MAE, and MAD are defined in Sec. \ref{sec: statistical and indicator}, while $\Bar{\chi}^2$ represents the mean reduced chi-square.}
    \label{table: number of eigenspectra}
    \begin{tabular}{cccccccc}
    \hline \hline
    Model & $m$ & GF & $\Bar{\Delta} z$ & MAE & MAD  & $\Bar{\chi}^2$ \\
     & &  & $/10{^-3}$  & $/10^{-3}$ & $/10^{-3}$ \\
    \hline
    $\text{Econder}_s$ & 3  & 0.9993 & 0.077 & 0.077 & 0.043 & 1.416 \\
    $\text{Econder}_s$ & 5  & 0.9995 & 0.077 & 0.077 & 0.042 & 1.311 \\
    $\text{Econder}_s$ & 10 & 0.9995 & 0.077 & 0.077 & 0.043 & 1.096\\
    
    \hline
    $\text{Econder}_g$ & 3  & 0.9791 & 1.510 & 2.240 & 0.035 & 1.610 \\
    $\text{Econder}_g$ & 5  & 0.9791 & 1.110 & 1.662 & 0.035 & 1.298 \\
    $\text{Econder}_g$ & 10 & 0.9757 & 1.311 & 2.005 & 0.035 & 1.211\\

    \hline
    $\text{Econder}_q$ & 3  & 0.9115 & 17.50 & 46.38 & 0.230 & 1.507 \\
    $\text{Econder}_q$ & 5  & 0.9107 & 16.91 & 43.51 & 0.236 & 1.465 \\
    $\text{Econder}_q$ & 10 & 0.9246 & 12.68 & 34.46 & 0.231 & 1.402\\
    \hline \hline
    \end{tabular}
\end{table}

In Sec. \ref{sec: data}, we have described the preprocessing step consisting of applying a fifth-order polynomial fitting, while in Sec. \ref{sec: models} we have adopted ten rest-frame eigenspectra from an Encoder to reconstruct the input spectrum, meaning only 16 parameters are required for its reconstruction. As both the preprocessing step and the number of eigenspectra are expected to affect the quality of the reconstructed spectrum, we want to determine the minimum representation needed for such a reconstruction in this section. To assess the impact of both the preprocessing and the number of eigenspectra, we employ a quadratic polynomial fitting in the spectrum preprocessing and then, with this new preprocessed spectrum, use 3, 5, and 10 eigenspectra, respectively.
Table \ref{table: number of eigenspectra} shows the model's performance after each model is trained on 20,000 spectra.
%with varying numbers of eigenspectra, with 

As the number of eigenspectra increases, the average minimum chi-squared values, $\Bar{\chi}^2$, decrease, as expected. The case with the fewest eigenspectra, shown in Table \ref{table: number of eigenspectra}, where a quadratic polynomial fitting and three rest-frame eigenspectra are used, requires only six modeling parameters for each input spectrum. This configuration yields a GF of 99.93\%, 97.91\%, and 91.15\% and an average minimum chi-square of 1.416, 1.610, and 1.507 for STAR, GALAXY, and QSO, respectively.

Reducing the number of modeling parameters is intended to find an optimal balance between fitting ability, generalization, and interpretability of the model. Using fewer parameters enhances the generalization and interpretability of the model. Our test demonstrates that the spectrum can be effectively modeled with only six parameters, achieving a GF above 90\% and an average minimum chi-square below $\sim 1.6$. However, to achieve optimal performance, the number of eigenspectra should be $\ge 5$. 
The chi-square suggests that using more eigenspectra leads to better spectrum fitting. However, for redshift estimation, a higher GF is obtained with the spectra of galaxy $m=5$ than with $m=10$.

Finally, we can evaluate the impact of the preprocessing with different degrees of polynomial fitting by directly comparing the quadratic polynomial case with $m=10$ eigenspectra (see Table \ref{table: number of eigenspectra}) with the fifth-polynomial fitting (Encoder in Table \ref{table: chi2 and D}), where we have assumed $m=10$. From this comparison, we see that the fifth-order polynomial preprocessing gives a lower average $\Bar{\chi}^2$, with respect to the quadratic one, although the improvement is small.

\subsection{Anomalies}
\label{sec: anomaly}

\begin{figure}
    \hspace{-0.5cm}
    \includegraphics[width=0.5\textwidth]{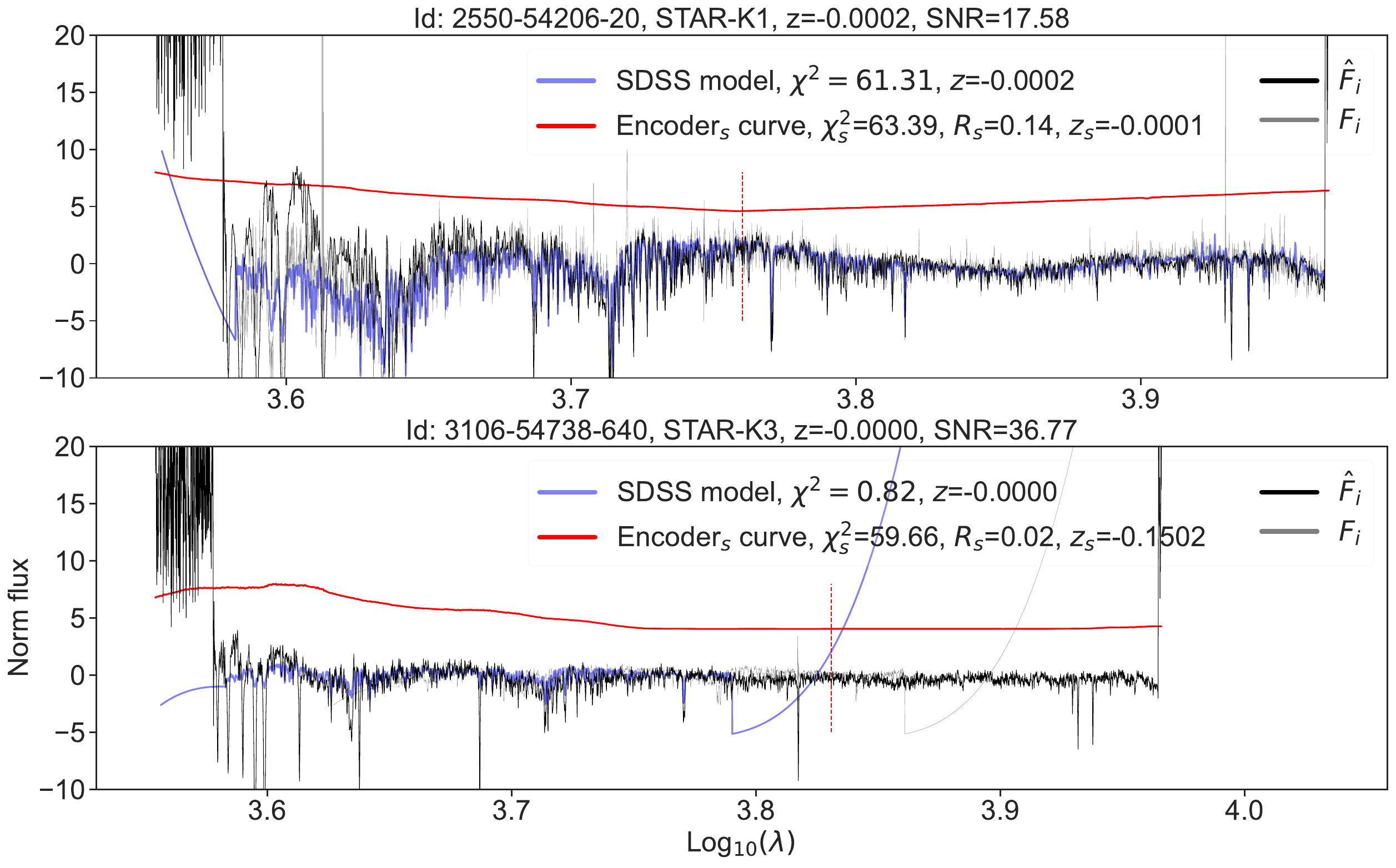}
    %~\\
    \hspace{0.5cm}

     \hspace{-0.5cm}
    \includegraphics[width=0.51\textwidth]{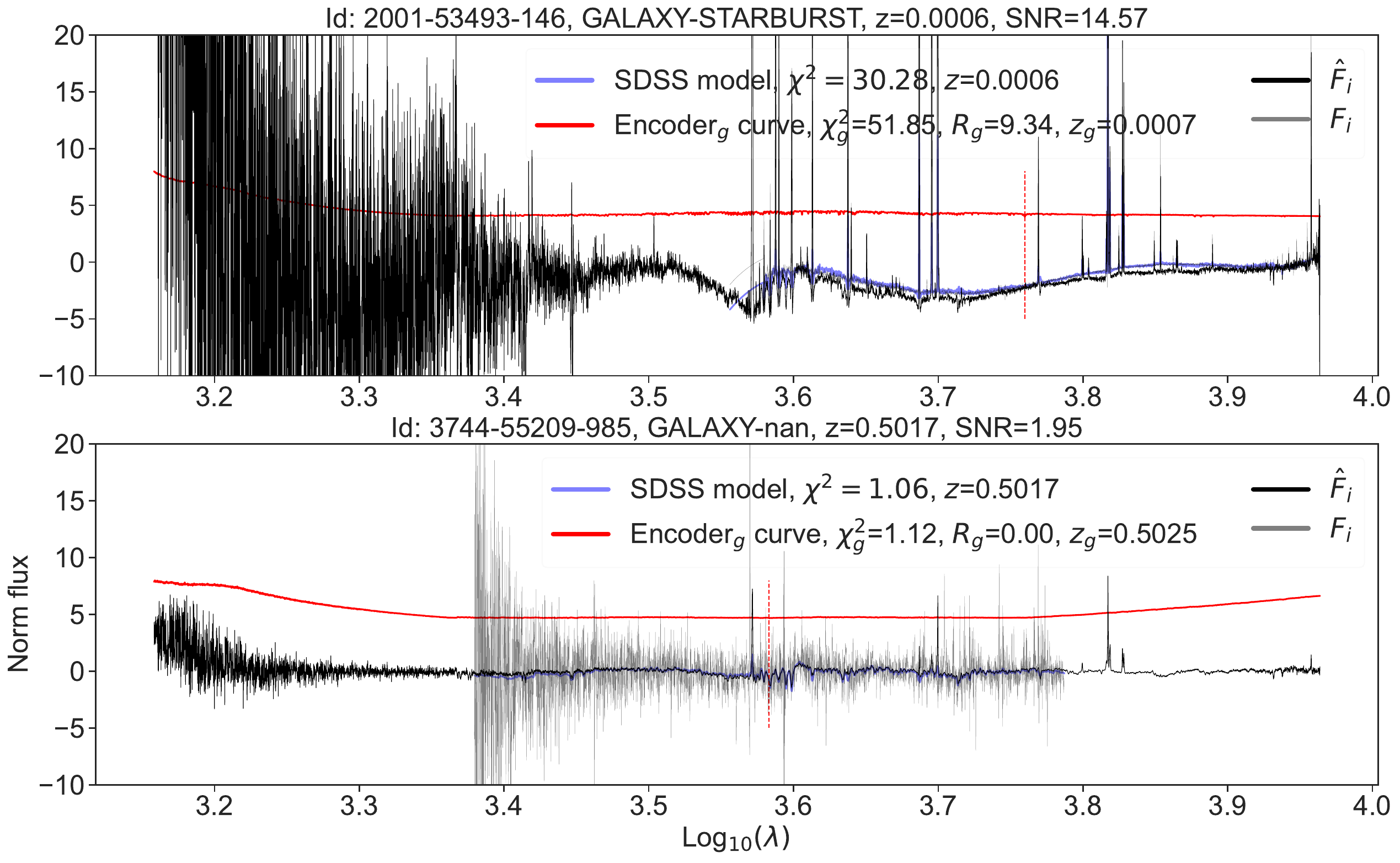}
    %~\\
    \hspace{0.5cm}
    
    \hspace{-0.5cm}
    \includegraphics[width=0.51\textwidth]{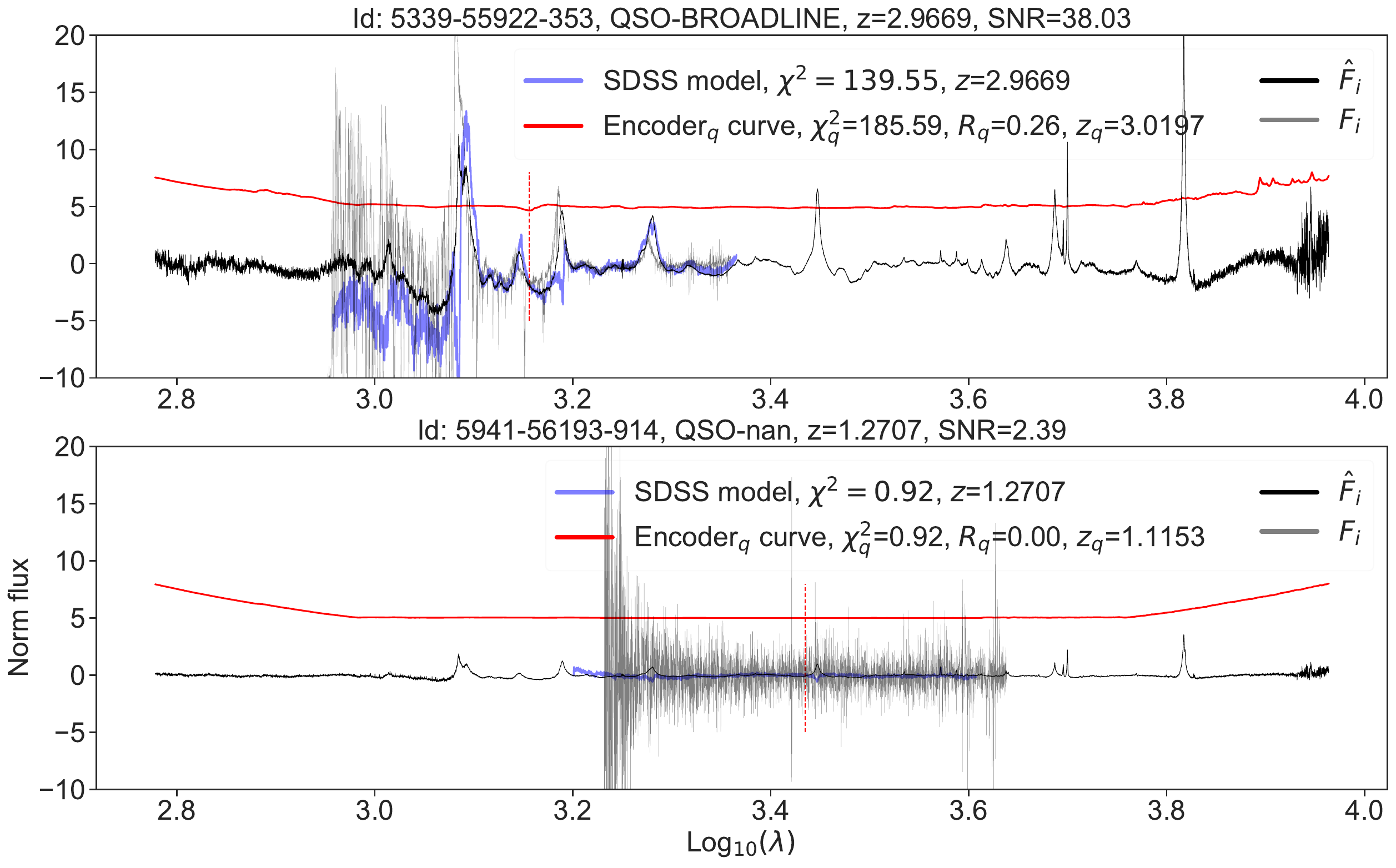}
    \caption{Examples of maximum chi-square and minimum robustness cases in the test spectra. From top to bottom, the cases include STAR, GALAXY, and QSO spectra. The blue dashed line represents the SDSS model, while the other labels are identical to Fig. \ref{fig: normal QSO}.} 
    \label{fig: anomaly case}
\end{figure}

Spectrum anomalies include a wide variety of deviations from some regular pattern expected for a given class of astronomical sources. Defects in the data also fall into this category, and their identification is part of the quality check of survey products. More interestingly, though, are anomalies having physical meaning, as these might contain information on missing knowledge. For instance, to be identified as anomalies, these deviations have to be defined with respect to a ``model''. Thus, if the model is incomplete or failing, anomalies allow us to identify the missing physics behind the model. However, there might be anomalies exceeding the features expected in a given class, for example, if these deviations are peculiar to a new class or unaccounted phenomena. In all these cases, the identification of anomalies can bring important information for new discoveries. 

To schematize, we can tentatively group the spectra anomalies into three kinds: 1) data defects, 2) anomalies from incomplete models, and 3) extreme anomalies exceeding well-understood models.
%(including unknowns).
To quantitatively identify the anomalies of these three kinds, we can treat them as outliers in diagnostics that measure the ``distance" of a given spectrum from standard features. A straightforward way is to use the $\chi^2$ and $R$. E.g., spectra with a large $\chi^2$ for all classes would imply a very uncertain classification, while a poor Robustness (i.e., $R\sim 0$) might indicate the presence of multiple equivalent minima for the redshift solution. Some examples of both large $\chi^2$ and minimum $R$ obtained from the test sample are reported in Fig. \ref{fig: anomaly case}, which eventually represent the catastrophic redshift predictions and misclassifications seen in Fig. \ref{fig: Encoder redshift}. 

%Here, we need to comment that by anomalies, we can certainly classify spectra that deviate from standard spectra in the three classes because they represent peculiar objects. However, we should also consider that in this automated process, there might still be an incomplete representation of the spectra derived in the training set. This lack of information can still produce large chi2 or lower R, hence anomalies.

%To be finished

%Below, we consider some clear cases for these conditions on the $\chi^2$ and $R$:
Among these cases, we can recognize:
\begin{enumerate}
    \item Low SNR causing redshift degeneracy. When there are multiple close local minima or even no significant peaks in the chi-squared curve, the $R$ is small, and the best-fit redshift becomes uncertain. This is demonstrated in rows 4 and 6 of Fig. \ref{fig: anomaly case}, where $R=0.00$. In these cases, the low SNR blurs the distinct spectral line features for confident redshift determination. This case cannot be considered an anomaly of any of the three kinds above.

    \item Large $\chi^2$ caused by a very intense emission line or a defective spectrum. Since chi-square fitting takes into account the entire input wavelength range, mismatches may occur when dealing with very strong and narrow emission lines, resulting in a large chi-squared value. These cases are shown in rows 1 and 3 of Fig. \ref{fig: anomaly case}.
    Similarly, spectra that are partially defective can lead to poor fits and higher chi-squared values. This case is shown in row 2 of Fig. \ref{fig: anomaly case}. These cases are consistent with anomalies of the first and second kinds illustrated above. In particular, for the first kind, we can check if, by adding spectra with strong emission lines, we could correctly recover these spectra.

    \item Large chi-square caused by Broad Absorption Line (BAL) QSO and Lyman $\alpha$ forest. The high-redshift QSO spectrum is usually imprinted by Lyman $\alpha$ absorption, which makes them irregular and challenging to fit with eigenspectra.
    Additionally, the BAL region is also not easily perfectly fitted by eigenspectra, as the discrepancy between the SDSS model and Encoder reconstruction is evident in row 5 of Fig. \ref{fig: anomaly case}. This is again the case of the second kind, as we acknowledge that there is an insufficient number of BAL QSO in training samples, preventing the model from capturing its features.
    \end{enumerate}

As anticipated, some anomalous cases can become normal if the reconstruction model is also able to capture peculiar features of known objects.
%y represent a sub-class that has not been preventively accounted. In this respect, the presence or recurrent properties in a given subset of anomalous cases can help identify this subclass and elaborate strategies to be classified.  
For instance, objects with intense emission lines in stars or galaxies can produce large $\chi^2$ that can be corrected by fitting individual lines separately or using a specialized training sample. Additionally, spectrum preprocessing can include some procedures to cut cataloged strong emission lines.
\cite{2017MNRAS.465.4530B} demonstrated various anomalies by comparing synthetic spectra with observed spectra using Random Forests. Their more sophisticated spectral analysis revealed that the primary cause of these anomalies is unusual line features, such as BPT (Baldwin, Phillips \& Terlevich diagrams; \citealt{1981PASP...93....5B, 1987ApJS...63..295V}) outliers or strong, weak, and broad emission lines. So, analyzing the residuals is crucial to detecting anomalies in scientific applications. This would be a reason for improving the reconstruction of peculiar cases or for identifying some missing physics in the emission line process. 
Additionally, the asymmetry of broad emission lines may pose challenges for precise redshift estimation. However, we believe it will not affect the coarse classification results. In our method, such asymmetries can be defined as anomalies for further analysis.

More difficult is to predict situations where anomalies can bring new discoveries. Some textbook examples are cases where the residual spectrum with inconsistent redshifted emission or absorption lines may indicate background emission from strong gravitational lensing systems (e.g., \citealt{2006ApJ...638..703B}) or intergalactic chemical absorption from foreground structures (e.g., \citealt{2010A&A...510A..54W}). 
%However, further detailed analysis of such anomalies is beyond the scope of this paper.
%Overall, the discussion above shows the importance to, first, spot possible anomalous cases in large pile of data in an automatic way, and, second, eventually classify them to optimise the outcome in terms of quality check and science exploitation.

To conclude, in this section, we have tried to give a few examples, although far from complete, illustrating the potential offered by the usage of even simple diagnostics to identify outliers that might contain important information to consider for both quality checks and science.

\section{conclusion}
\label{sec: conclusion}

In this work, we have presented two generative networks designed to reconstruct the rest-frame spectrum from processed observed spectra from SDSS (taken as a reference dataset): an Encoder and an UNet model. 
The Encoder model generates complete spectrum modeling parameters, including rest-frame eigenspectra derived from the training spectra and corresponding coefficients. In our implementation, we used 16 parameters for each spectrum: 10 eigenspectra and a fifth-order polynomial for continuous fitting of the input spectrum. The rest-frame eigenspectra produced by the Encoder reveal absorption and emission line features effectively captured by the network. This approach enhances the interpretability of Encoder in spectrum modeling compared to the end-to-end UNet.

Both models are built-in stand-alone algorithms, named Galaxy Spectra Network - III (GaSNet-III, after previous developments in GasNet-I and GasNet-II), performing 1) spectrum classification, 2) redshift estimate, and 3) anomaly detection. Indeed, the reconstructed rest-frame spectra have been used for a coarse spectral classification (in STAR, GALAXY, and QSO-types), while the redshift prediction is obtained by ``moving'' the rest-frame reconstructed spectrum to match the input spectrum. For this latter task, for the first time, we can demonstrate that a DL method can reach the same level of accuracy as classical methods based on cross-correlation or template fitting but with a computational speed that is about three orders of magnitude faster. Besides this clear advantage with respect to standard tools, GaSNet-III makes use of reconstructed ``model'' spectra to recognize strongly deviating spectra, possibly identifying anomalies.

This has been made possible by scanning the entire parameter space of spectral type (S/G/Q) and redshift with the computational efficiency of DL. We have generated $\chi^2$ curves in the whole parameter space. This has enabled us to identify the global $\chi^2$ minimum and assess the robustness, $R$, within these curves. 

The average classification accuracy is 98\% for STAR-GALAXY-QSO classification, and 99.9\%, 97.4\%, and 98.6\% for stars, galaxies, and QSOs, respectively. The fraction of estimated redshifts with a relative error below the required thresholds of $\sim$$10^{-4}$ for stars and galaxies, and $\sim$$10^{-3}$ for QSOs (defined as a good fraction, GF) is approximately 99.9\%, 97.8\%, 92.8\% for STAR, GALAXY, and QSO spectra, respectively, using the SDSS pipeline classification and redshifts as ground truth.
%The average classification accuracy is 98\% for STAR-GALAXY-QSO classification and 99\%, 98\%, while the fraction of estimated redshifts below the required relative error of $\sim$$10^{-4}$ for stars and galaxies and $\sim$$10^{-3}$ for quasars (defined as good-fraction, GF) is $\sim$93\% for STAR, GALAXY, and QSO spectra, respectively, using the SDSS pipeline classification and redshifts as ground truth. 
These results are compared with standard methods like QXP++ {(Davies et al. in prep)}, Redmonster \citep{2016AJ....152..205H}, and pPXF \citep{2023MNRAS.526.3273C}, and only Redmonster has shown performances slightly better than GaSNet-III.

The enhanced performance of GaSNet-III resides in the well-designed Encoder network architecture, which is specifically sensitive to the redshift translation symmetry in logarithmic wavelength space and the statistical similarities across observed spectra. By subtracting the continuum using a polynomial fit, we directed the network to model absorption and emission features in the residuals while capturing statistical similarities via a set of rest-frame eigenspectra learned from the training set. 

The performances of the two networks, though, are comparable but with some differences: the Encoder model achieves a lower average $\chi^2$ in STAR and GALAXY spectrum modeling and a comparable chi-square in QSO modeling when compared to the SDSS pipeline, while our UNet achieves a lower average $\chi^2$ on all types of spectra. 

As anticipated, the robustness and minimum $\chi^2$ values serve as indicators of confidence in best-fit estimates, but they have been used as a metric to identify potential outliers or anomalies.
The reconstructed rest-frame spectra from pre-trained models exhibit reduced noise levels. By calculating $\chi^2$ curves between the reconstructed and input spectra across different pre-trained models, we determine not only the best-fit class and redshift but also the identity of the potential anomalies. Anomalies primarily include galaxies with extremely strong narrow emission lines, high-redshift BAL QSO spectra, and defective spectra, which can be indicated by robustness and best-fit chi-squared values. Setting a lower threshold of $R \ge 0.1$ can effectively remove most cases of catastrophic redshift prediction (see Fig. \ref{fig: D_vs_GF}).

%We also explored a minimal representation for the input spectrum that uses only six parameters: three eigenspectra and a quadratic polynomial for continuum fitting. This configuration successfully models most spectra with a GF exceeding 90\%. 

Our new tool achieves the same level of modeling accuracy as traditional methods, such as the SDSS pipeline, while providing a new, DL-based solution for accurate and high-efficiency spectrum modeling, which is particularly valuable for handling massive data volumes from future spectroscopic sky surveys.
In future work, we plan to incorporate velocity dispersion as an additional parameter in the Encoder model to increase its interpretability.

\section*{Acknowledgements}
We thank Dr. Domitilla De Martino for her valuable comments.
FZ acknowledges the support of the China Scholarship Council (grant n. 202306380249). NRN acknowledges support from the Guangdong Science Foundation grant (ID: 2022A1515012251). CH's work is funded by the Volkswagen Foundation. CH also acknowledges support from the Deutsche Forschungsgemeinschaft (DFG, German Research Foundation) under Germany's Excellence Strategy EXC 2181/1 - 390900948 (the Heidelberg STRUCTURES Excellence Cluster). R.D. gratefully acknowledges support by the ANID BASAL project FB210003. BFR acknowledges the Polish LSST/Rubin grant: MNiSW grant DIR/WK/2018/12. CS acknowledges support from the Agencia Nacional de Investigaci\'on y Desarrollo (ANID) through Basal project FB210003. AV acknowledges the support from the SNF 200020\_175751 and 200020\_207379 "Cosmology with 3D Maps of the Universe" research grant. RJA was supported by FONDECYT grant number 1231718 and by the ANID BASAL project FB210003.

\section*{DATA AVAILABILITY}
The code and example data are available at the following GitHub link: \url{https://github.com/Fucheng-Zhong/GaSNet-III}.

\bibliography{references}
\bibliographystyle{mnras}
%%%%%%%%%%%%%%%%%%%%%%%%%%%%%%%%%%%%%%%%%%%%%%%%%%
%%%%%%%%%%%%%%%%% APPENDICES %%%%%%%%%%%%%%%%%%%%%
%\newpage
%\clearpage
\appendix

\section{Visualization}
\label{sec: outliers}
\begin{figure*}
    \centering
    \includegraphics[width=0.9\textwidth]{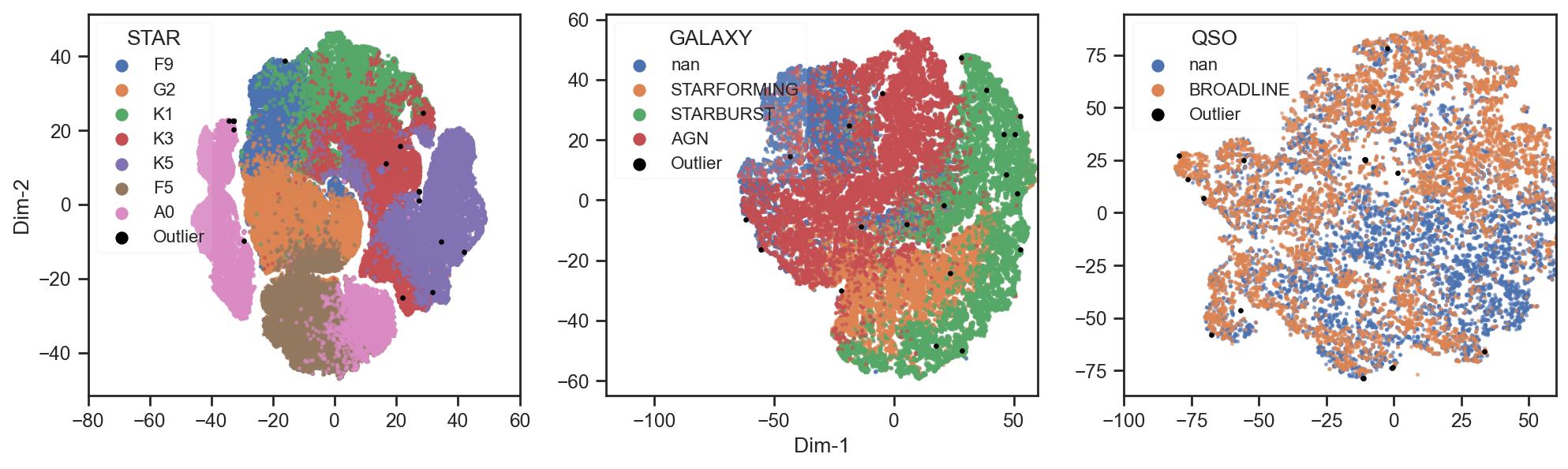}
    \caption{Two-dimensional t-SNE projections of the coefficients with subclass labels color-coded. The outliers found by LOF are labeled in black.} 
    \label{fig: tSNE_visualization_subclass2}
\end{figure*}

In this section, we visualize and analyze the distribution of coefficients. Typically, different subclasses occupy distinct regions within the coefficient space, suggesting that unsupervised clustering methods could potentially be used to classify these subclasses effectively.
We project the ten coefficients into a two-dimensional representation using t-SNE \citep{van2008visualizing}. In Fig. \ref{fig: tSNE_visualization_subclass2}, the projection is labeled by subclass. The majority of STAR and GALAXY subclasses are well-separated, indicating a clear distinction, while the QSO subclasses show significant overlap, with many points clustering together. This suggests that further refinement may be needed to differentiate QSO subclasses.

This analysis demonstrates that an unsupervised classification approach based on clustering methods can effectively classify STAR and GALAXY subclasses. However, for QSO, the distinction of broad-line subclasses is less pronounced, indicating that coefficient clustering alone may not be sufficient for accurate subclass classification of QSOs.

Additionally, to identify outliers in coefficient space based on cluster density, we applied the Local Outlier Factor (LOF; see \citealt{alghushairy2020review}), also known as the anomaly score, to mark outliers on the t-SNE representation. However, the drawback is its computational complexity of $\mathcal{O}(n^2)$. The results show that LOF-detected anomalies generally lie within the edges of the primary cluster areas.

\section{Impact of the training set sizes}
\label{Appendix: data number}

\begin{table}
    \centering
    \caption{The performance of different training set sizes. Num represents the size of the training set.}
    \label{table: number of training data}
    \begin{tabular}{ccccccc}
    \hline \hline
    Model & Num & GF & $\Bar{\Delta} z$ & MAE & MAD  & $\Bar{\chi}^2$ \\
     & $10^3$ & \% & $/10^{-3}$  & $/10^{-3}$ & $/10^{-3}$ & \\
    \hline
    $\textbf{Econder}_s$ & 5  & 99.93 & 0.079 & 0.079 & 0.043 & 1.306 \\
    $\textbf{Econder}_s$ & 10 & 99.94 & 0.079 & 0.079 & 0.045 & 1.144 \\
    $\textbf{Econder}_s$ & 20 & 99.95 & 0.077 & 0.077 & 0.043 & 1.119 \\

    \hline
    $\textbf{Econder}_g$ & 5  & 94.59 & 2.814 & 4.391 & 0.037 & 1.592 \\
    $\textbf{Econder}_g$ & 10 & 97.26 & 1.148 & 1.786 & 0.036 & 1.296 \\
    $\textbf{Econder}_g$ & 20 & 97.49 & 1.397 & 2.132 & 0.035 & 1.219 \\
    
    \hline
    $\textbf{Econder}_q$ & 5  & 90.74 & 17.67 & 51.15 & 0.208 & 1.552 \\
    $\textbf{Econder}_q$ & 10 & 91.85 & 14.90 & 43.08 & 0.234 & 1.435 \\
    $\textbf{Econder}_q$ & 20 & 92.02 & 15.47 & 40.21 & 0.245 & 1.432 \\
    
    \hline \hline
    \end{tabular}
\end{table}

The number of well-modeled or well-calibrated observed spectra with accurate redshifts is often limited and not always as abundant as desired for model training. In this section, we investigate the performance of the Encoder model with varying training set sizes, selecting sets of 5,000, 10,000, and 20,000 spectra for each class. For each set, we apply a quadratic polynomial fitting in spectrum preprocessing and utilize ten eigenspectra in the model.

Table \ref{table: number of training data} presents the performance of models across different training set sizes. The results indicate that the model reaches near-optimal performance with training sets of 10,000 spectra or more. Increasing the amount of training data beyond this point offers only tiny gains, likely due to the limited diversity within the training samples, as the network has already captured most of the essential spectral features.

To further enhance performance, selecting training spectra that are representative and diverse could be more effective than randomly selecting samples from the survey. By pre-selecting spectra that cover a variety of observed spectra, we can probably improve the model performance from a broader and more representative dataset.

\section{Examples of Encoder modeling}
\label{Appendix: Examples Encoder modeling}
\begin{figure*}
    \centering
    \includegraphics[width=0.75\textwidth]{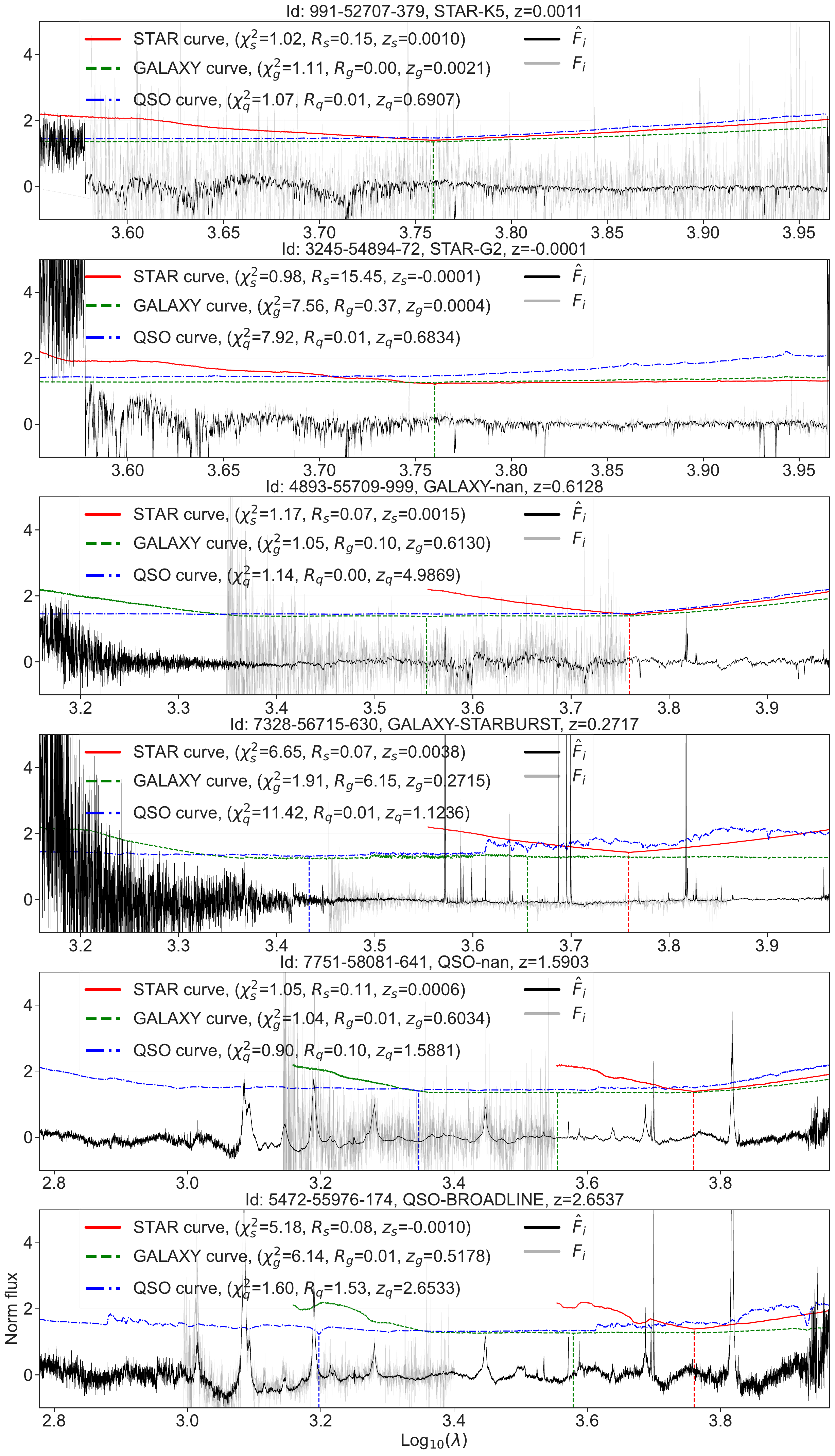}
    \caption{Examples of input and reconstructed STAR, GALAXY, and QSO spectra. The labels are identical to Fig. \ref{fig: normal QSO}.}
    \label{fig: normal reconstructed}
\end{figure*}

We present six randomly selected normal spectra reconstructed by Encoder in Fig. \ref{fig: normal reconstructed}.

\end{document}